# Tweet Influence on Market Trends: Analyzing the Impact of Social Media Sentiment on Biotech Stocks

C. Sarai R. Avila[1]

[1]College of Science and Enginering, School of Mathematics and Statistics, Univeristy of

Glasgow.

June 2023

**ABSTRACT.** This study investigates the relationship between tweet sentiment across diverse categories: news, company opinions, CEO opinions, competitor opinions, and stock market behavior in the biotechnology sector, with a focus on understanding the impact of social media discourse on investor sentiment and decision-making processes. We analyzed historical stock market data for ten of the largest and most influential pharmaceutical companies alongside Twitter data related to COVID-19, vaccines, the companies, and their respective CEOs. Using VADER sentiment analysis, we examined the sentiment scores of tweets and assessed their relationships with stock market performance. We employed ARIMA (AutoRegressive Integrated Moving Average) and VAR (Vector AutoRegression) models to forecast stock market performance, incorporating sentiment covariates to improve predictions. Our findings revealed a complex interplay between tweet sentiment, news, biotech companies, their CEOs, and stock market performance, emphasizing the importance of considering diverse factors when modeling and predicting stock prices. This study provides valuable insights into the influence of social media on the financial sector and lays a foundation for future research aimed at refining stock price prediction models.

**EXECUTIVE SUMMARY**

The rapidly evolving landscape of social media has significantly impacted the financial sector, with platforms like Twitter becoming influential in shaping public opinion and disseminating financial information. This study aims to investigate the relationship between various tweet categories: news, company opinions, CEO opinions, competitor opinions, and stock market behavior, with a focus on understanding how tweet sentiment impacts investor sentiment and decision-making processes.

The growing importance of social media in finance is evidenced by the democratization of financial news access and the rise of online investor communities. Twitter has emerged as a vital platform for financial information, with its real-time nature and widespread adoption by various stakeholders in the financial sector. Previous research has shown a correlation between Twitter sentiment and stock market behavior (Ranco et al. 2015; Mendoza-Urdiales et al. 2022; Katsafados, Nikoloutsopoulos, and Leledakis 2023; Das et al. 2018; Pagolu et al. 2016). However, there remains a gap in understanding the broader impact of diverse tweet categories on the stock market.

This study aims to address this gap and provide valuable insights into the intricate relationship between Twitter discourse and stock market behavior. By examining the impact of distinct tweet types, the research will shed light on the mechanisms through which Twitter affects investor sentiment and decision-making, ultimately influencing market outcomes. A key contribution of this study is the potential improvement of stock market predictions by incorporating opinions from various tweet categories into predictive models. The insights gained from this research can inform strategies for market participants, regulators, and policymakers to better navigate and respond to the rapidly evolving landscape of social media and its impact on the financial sector.

To investigate the relationship between Twitter discourse and stock market behavior, this study focuses on the biotechnology sector, analyzing ten of the largest and most influential pharmaceutical companies in the industry. Historical stock market data for these companies was obtained from the Yahoo Finance API in Python, capturing daily and hourly data points from February 1, 2023, to March 19, 2023.

In parallel, Twitter data related to COVID-19, vaccines, the ten biotech companies, and their respective CEOs were collected employing the Tweepy library. Due to Twitter's API limitations, tweets were collected every three to five days between February 1, 2023, and March 19, 2023, ensuring complete coverage of the relevant period. Data collection occurred during stock market trading hours, allowing for the use of Twitter data to model stock market performance. Search queries included terms: COVID-19, vaccines, company names, and CEO names.

Our analysis of stock market data for ten biotech companies revealed increased volatility in after-hours trading. We also observed the volume of tweets mentioning these companies, their CEOs, and news related to COVID-19 and vaccines, which provided insight into public engagement and discussions surrounding these entities.

Using VADER sentiment analysis, we examined the sentiment scores of tweets related to these companies, their CEOs, and news events. We found that tweets mostly expressed mild positive or negative emotions or predominantly neutral sentiments, with some specific companies and CEOs attracting more positive sentiments.

We then assessed the relationships between sentiment scores and stock market performance. Notably, we found a positive correlation between the sentiment scores of certain companies and their competitors, highlighting the importance of analyzing competitor sentiment when trying to understand factors that could affect stock prices.

We utilized the ARIMA (AutoRegressive Integrated Moving Average) model to forecast stock market performance and found that incorporating sentiment covariates into the model led to improved predictions. The vaccine-related covariate generally performed better as polarized opinions on vaccines impact investor sentiment, but it was crucial to consider other covariates and their combinations to achieve the most accurate predictions for each specific company.

Lastly, we employed the VAR (Vector AutoRegression) model to predict stock prices while accounting for the influence of various sentiment factors. In some cases, incorporating the sentiment covariates significantly improved the prediction accuracy, emphasizing the importance of considering diverse factors when modeling and predicting stock prices.

In conclusion, our study successfully investigated the relationship between news sentiment, biotech companies, their CEOs, and stock market performance. By combining time series forecasting techniques and sentiment analysis, we provided valuable insights into the complex interplay of these factors. This study emphasizes the importance of considering various factors when modeling and predicting stock prices and provides a foundation for future research incorporating diverse factors for more accurate stock price predictions.

Potential directions for future work include expanding the scope of the study, delving deeper into the impact of specific events or announcements, obtaining social media data from other sources, exploring other sentiment analysis techniques, including additional features and covariates, comparing the performance of different time series forecasting techniques, and developing real-time sentiment analysis and stock price prediction systems. By pursuing these research directions, we can continue to advance our understanding of the relationship between news sentiment, biotech companies, their CEOs, and stock market performance, and develop more effective and accurate stock price prediction models.

1. **INTRODUCTION**

In recent years, social media has become a powerful force in shaping public opinion and disseminating information across various sectors, including finance. Platforms like Twitter have significantly changed how investors and market participants access and process financial news, complementing and sometimes replacing traditional sources such as newspapers and television broadcasts (Dwivedi et al. 2021).

The stock market, as a complex and dynamic system, is influenced by numerous factors, such as economic indicators, corporate performance, and global events. Investor sentiment and market psychology, driven by news and opinions shared on social media, play a crucial role in shaping stock prices and market trends. Thus, understanding the relationship between social media, particularly Twitter, and the stock market is essential in today's financial landscape (Zhang et al. 2022).

Previous studies have primarily focused on company opinions on Twitter, (Ranco et al. 2015; Mendoza-Urdiales et al. 2022; Katsafados, Nikoloutsopoulos, and Leledakis 2023; Das et al. 2018; Pagolu et al. 2016), leaving a gap in understanding the broader impact of diverse tweet categories, such as news, CEOs opinions, and competitor opinions, on the stock market. This study aims to address this gap by systematically investigating the relationship between various tweet types and stock market behavior. The research questions include examining the extent to which news, company, and competitor opinions shared on Twitter influence stock market trends, as well as exploring how tweet sentiment impacts investor sentiment and decision-making processes in the stock market.

A key contribution of this study is the potential improvement of stock market predictions by incorporating opinions from various tweet categories into predictive models. More accurate and reliable predictive models can benefit market participants by helping them make better-informed decisions, fostering more efficient and transparent financial markets. Furthermore, this research can provide a foundation for future studies on the influence of other social media platforms on financial markets.

The significance of this study lies in its potential to enhance our understanding of the intricate relationship between Twitter discourse and stock market behavior, providing valuable insights for various stakeholders in the financial sector. By examining the impact of distinct tweet types, the research will shed light on the mechanisms through which Twitter affects investor sentiment and decision-making, ultimately influencing market outcomes. The insights gained from this research can inform strategies for market participants, regulators, and policymakers to better navigate and respond to the rapidly evolving landscape of social media and its impact on the financial sector.

## 2. LITERATURE REVIEW

### 2.1 Social Media and Finance

The financial sector has experienced significant changes in recent years, with social media emerging as a powerful force in shaping the way market participants access and process information. The rise of social media has democratized access to financial news, analysis, and opinions, allowing individual investors to engage in real-time discussions and make investment decisions based on up-to-date information (Biancotti and Ciocca n.d.). Moreover, social media has enabled companies to directly communicate with their stakeholders, providing transparency and fostering investor confidence (Dwivedi et al. 2021).

The growing influence of social media in the financial sector can also be observed in the emergence of online communities and forums, where investors gather to share trading strategies, stock tips, and market analyses. These online interactions can have a significant impact on stock prices and market trends, as evidenced by events like the GameStop short squeeze in early 2021("Implications of 'Short Squeezes' on Financial Market Efficiency" n.d.).

As social media continues to gain prominence in the financial world, market participants are increasingly recognizing the need to monitor and analyze social media data to make informed decisions and anticipate market trends. This shift has led to the development of new tools and techniques, such as sentiment analysis and natural language processing being applied in the finance field, that enable the extraction of valuable insights from social media data to inform investment strategies (Kapoor et al. 2018).

The impact of social media on financial markets can be observed across various platforms, each with its unique features and user base. Twitter has emerged as a vital platform for financial information in the digital age, playing a pivotal role in shaping investor sentiment and market trends (Ranco et al. 2015; Das et al. 2018). Twitterreal-time nature and concise format make it an ideal medium for rapidly disseminating news and opinions, allowing market participants to react quickly to market-moving events(Das et al. 2018). Additionally, the platform's widespread adoption by investors, analysts, journalists, and company executives have fostered an environment where diverse perspectives and insights can be easily accessed and shared. The growing importance of Twitter in the financial landscape is evidenced by the fact that stock prices and market trends are often influenced by tweets from influential figures and organizations. For instance, tweets from CEOs, such as Elon Musk, have been known to cause significant fluctuations in stock prices (Ante 2023; Molla 2021). Moreover, the platform has enabled the rise of financial influencers, who leverage their expertise and online following to shape market sentiment and offer investment advice (Ante 2023).

Twitter serves as a hub for various types of financial information, including:

- News: Financial news organizations and journalists use Twitter to share breaking news, updates, and analyses. This real-time information enables investors to stay informed about market-moving events and make timely investment decisions.
- Company Opinions: Companies often share their perspectives on industry trends, product launches, financial results, and other market-related topics through their official Twitter accounts. These tweets provide investors with valuable insights into a company's performance, strategy, and outlook.
- CEO Opinions: CEOs and other high-ranking executives use Twitter as a platform to express their opinions on various topics, ranging from company performance and industry trends to broader economic and geopolitical issues. These tweets can have a significant impact on investor sentiment, as they often provide a unique perspective on a company's vision and strategic direction (Ante 2023; Molla 2021).

- Competitor Opinions: Twitter also serves as a platform for sharing opinions about competitors within an industry. The opinions of competitors can influence the company's investor sentiment and decision-making processes.

### 2.2 Previous Studies on Twitter and stock market behavior

The relationship between Twitter sentiment and stock market behavior has been extensively studied in recent years. Researchers have explored various aspects of this relationship, including the impact of tweet volume, sentiment polarity, and event-driven sentiment changes on stock prices and market trends.

One of the early studies by Bollen *et al.* demonstrated that the mood of Twitter users could predict stock market movements. Their research employed OpinionFinder and Google-Profile of Mood States to analyze the sentiment of tweets and found a correlation between the public mood on Twitter and the Dow Jones Industrial Average (DJIA) (Bollen, Mao, and Zeng 2011). Nguyen *et al.* studied the relationship between sentiment in tweets and stock market movements using a dataset of tweets mentioning S&P 500 companies. Their research employed machine learning techniques to predict stock market trends and showed that incorporating sentiment analysis significantly improved prediction accuracy (Nguyen, Shirai, and Velcin 2015). Ranco *et al.* analyzed the impact of tweet sentiment and volume on stock prices for 100 companies listed on the S&P 100 index *(Ranco et al. 2015)*. Their findings suggest that both sentiment and tweet volume could help predict stock price movements. They also highlighted the significance of event-driven sentiment changes in influencing stock market behavior. Mendoza-Urdiales *et al.* examined the role of Twitter sentiment in forecasting market volatility. By utilizing sentiment analysis and machine learning techniques, they found that incorporating sentiment information from tweets could improve the prediction of stock market volatility (Mendoza-Urdiales et al. 2022). Katsafados *et al.* investigated the relationship between Twitter sentiment and stock market trends for companies listed on the S&P 500 index (Katsafados, Nikoloutsopoulos, and Leledakis 2023). Their study combined sentiment analysis with a time-series analysis, revealing a significant relationship between Twitter sentiment and stock returns.

While these studies provide valuable insights into the relationship between Twitter sentiment and stock market behavior, there are still notable gaps in the literature. Many studies have focused primarily on company opinions, overlooking the impact of other tweet categories, such as news, CEO opinions, and competitor opinions. Furthermore, the generalizability of findings across different market segments or industries remains a concern.

### 2.3 Sentiment analysis for social media, focusing on Twitter

Sentiment analysis, also known as opinion mining or emotion AI, refers to the process of extracting subjective information, such as emotions, opinions, and attitudes, from textual data (Prabowo and Thelwall 2009; Hussein 2018; Mejova 2009; Medhat, Hassan, and Korashy 2014). In recent years, sentiment analysis has gained considerable attention in various domains, including finance, due to the increasing role of social media platforms, such as Twitter, in shaping public opinion and disseminating information.

Twitter, with its concise format and real-time updates, has emerged as a prominent source of information for various stakeholders, including investors, market analysts, and policymakers. As a result, researchers have increasingly focused on applying sentiment analysis techniques to Twitter data to better understand the impact of social media discourse on financial markets (Bollen, Mao, and Zeng 2011; Nguyen, Shirai, and Velcin 2015).

Various methods have been employed to perform sentiment analysis on Twitter data, ranging from traditional lexicon-based approaches to more advanced machine-learning techniques. Lexicon-based methods rely on pre-defined dictionaries of words associated with the positive, negative, or neutral sentiment, while machine learning techniques involve training algorithms on labeled datasets to classify tweets into sentiment categories (Pak and Paroubek 2010; Saif, He, and Alani n.d.)

Recent advances in natural language processing (NLP) and deep learning have further improved the accuracy and efficiency of sentiment analysis on social media platforms like Twitter. Techniques such as word embeddings and neural networks, including recurrent neural networks (RNNs) and transformer models, have been applied to capture complex semantic relationships and better classify tweets based on their sentiment (Babu and Kanaga 2022).

Despite these advancements, sentiment analysis on Twitter data presents unique challenges. The short and informal nature of tweets, along with the frequent use of abbreviations, slang, emojis, and sarcasm, can make it difficult for sentiment analysis algorithms to accurately capture the intended sentiment. Furthermore, context-specific nuances, such as financial jargon and industry-specific terminology, may not be adequately addressed by general sentiment analysis techniques, necessitating the development of domain-specific models (Loughran and Mcdonald 2011)

### 2.4 Predictions Models in Finance

Predictive models in finance have numerous applications, including stock market forecasting, portfolio management, and risk management. In stock market forecasting, predictive models are used to anticipate price movements and market trends, enabling investors to make informed decisions and maximize returns. Portfolio management involves the allocation of assets to optimize the risk-return profile of an investment portfolio. Predictive models, such as Markowitz's Modern Portfolio Theory, facilitate the construction of efficient portfolios that maximize expected returns for a given level of risk. Risk management entails the identification, assessment, and mitigation of financial risks. Predictive models, such as Value-at-Risk (VaR) and Conditional Value-at-Risk (CVaR), are used to estimate potential losses in a portfolio, allowing financial institutions and investors to implement risk mitigation strategies.

Numerous studies have explored and developed predictive models in finance, including those focused on stock market forecasting, portfolio optimization, and risk management. Some of the key studies in this domain include those by (Engle 1982) who introduced the Autoregressive Conditional Heteroskedasticity (ARCH) model for predicting stock market volatility, and Markowitz (1952), who developed the Modern Portfolio Theory (MPT) for optimizing investment portfolios. More recent studies have investigated the use of machine learning and artificial intelligence in finance, with researchers such as (Kimoto et al. 1990) and (Schumaker and Chen 2009) pioneering the application of neural networks and natural language processing to financial forecasting.

The rapid development of machine learning and artificial intelligence techniques has led to significant advancements in predictive models for finance. Researchers have started to explore the potential of deep learning, a subfield of machine learning, to improve financial forecasting accuracy. For instance, (Dixon, Klabjan, and Bang 2016) have employed deep learning techniques, such as Long Short-Term Memory (LSTM) networks and Convolutional Neural Networks (CNNs), to predict stock market trends. Additionally, ensemble methods, which combine the predictions of multiple models to improve accuracy, have gained popularity. (Ballings et al. 2015) applied ensemble techniques, such as bagging and boosting, to financial time-series forecasting and credit scoring, respectively, demonstrating the effectiveness of these approaches in various financial applications.

Predictive models in finance have traditionally relied on historical financial data and macroeconomic indicators. However, recent studies have started to explore the potential of alternative data sources, such as social media sentiment, news articles, and satellite images, for financial forecasting. As mentioned earlier, studies by (Bollen, Mao, and Zeng 2011; Nguyen, Shirai, and Velcin 2015) have investigated the relationship between Twitter sentiment and stock market trends, while (Paul C. Tetlock 2007) analyzed the impact of news sentiment on stock prices.

Despite the advancements and promising applications, predictive models in finance still face several challenges. One significant challenge is the inherent complexity and non-stationarity of financial markets, which can make accurate forecasting difficult. Another challenge is the high dimensionality of financial data, which may lead to overfitting and reduce the generalizability of models. Moreover, financial data can be noisy, making it challenging to extract meaningful patterns and signals. Data preprocessing techniques, such as feature selection, feature extraction, and data cleaning, are essential to address this issue but can be time-consuming and complex. Finally, the lack of transparency and interpretability in certain machine learning models, such as deep learning, can make it difficult for practitioners to trust and adopt these models in their decision-making processes.

### 3. DATA

#### 3.1 Acquisition of Historical Stock Market Data of the 10 studied biotech companies

In order to address our research question concerning the stock market, we chose to concentrate on the biotechnology sector. This decision was informed by the substantial impact and relevance of the biotech industry in recent years given the world pandemic of COVID-19, as well as its potential for future growth and innovation. The study focuses on ten of the largest and most influential pharmaceutical companies in the biotech industry, which include Johnson & Johnson, Eli Lilly, Novo Nordisk, AbbVie, Merck, Pfizer, Roche, AstraZeneca, Novartis, and Moderna (Sagonowsky et al. 2022).

Historical stock market data for these ten pharmaceutical companies, all listed on the New York Stock Exchange (NYSE), were obtained using the Yahoo Finance API in Python "yfinance". Data collection occurred from February 1, 2023, to March 19, 2023, capturing data points daily and hourly during NYSE trading hours. Although NYSE trading typically runs from 9:30 AM to 4:00 PM, data collection for this study continued until 3:30 PM, to respect the hourly collection. The data acquired encompassed opening and closing prices, high and low prices, adjusted closing prices, and the volume of shares traded for each day within the specified period.

List of companies that stock market data were collected:

- Johnson & Johnson

- Eli Lilly

- Novo Nordisk

- AbbVie

- Merck

- Pfizer

- Roche

- AstraZeneca

- Novartis

- Moderna

### 3.2 Collection of Twitter Data on News, Companies, and CEOs

To validate our hypothesis that news, companies, and CEOs affect the stock market, we collected tweets related to COVID-19, vaccines, the ten biotech companies mentioned in Section 3.1, and their respective CEOs. The CEOs included in the study are Joaquin Duato (Johnson & Johnson), David A. Ricks (Eli Lilly), Lars Fruergaard Jørgensen (Novo Nordisk), Richard A. Gonzalez (AbbVie), Kenneth C. Frazier (Merck), Albert Bourla (Pfizer), Severin Schwan (Roche), Pascal Soriot (AstraZeneca), Vas Narasimhan (Novartis), and Stéphane Bancel (Moderna).

Due to Twitter's API limitations, which only allow the retrieval of tweets from the past seven days, we collected tweets every three to five days between February 1, 2023, and March 19, 2023, to ensure complete coverage of the relevant period. This required diligent tracking and scheduling of data collection to avoid missing any crucial information.

Tweets were collected during the hours of 8:30 AM to 3:30 PM New York Time, aligning with the stock market trading hours. This allowed us to use Twitter data to model stock market performance, such that data collected from 8:30 AM to 9:30 AM could be used to predict the opening stock prices at 9:30 AM, and the same idea apply for the rest of the trading hours.

To collect the data, we used a Python script that employed the Tweepy library. We fetched tweets based on specific search queries related to COVID-19, vaccines, the ten biotech companies, and their respective CEOs. The search queries included "COVID," "vaccine," company names (e.g., "Johnson & Johnson," "Eli Lilly," etc.), and CEO names (e.g., "Joaquin Duato," "David A. Ricks," etc.), given a total number of 12 search queries. The script allowed us to search for tweets within a specified date and time range, ensuring that we collected data corresponding to the same period as the stock market data.

List of search queries used to collect tweet data:

1. COVID

2. Vaccine

3. Johnson & Johnson

4. Eli Lilly

5. Novo Nordisk

6. AbbVie

7. Merck

8. Pfizer

9. Roche

10. AstraZeneca

11. Novartis

12. Moderna

13. Joaquin Duato

14. David A. Ricks

15. Lars Fruergaard Jørgensen

16. Richard A. Gonzalez

17. Kenneth C. Frazier

18. Albert Bourla

19. Severin Schwan

20. Pascal Soriot

21. Vas Narasimhan

22. Stéphane Bancel

The output text files include columns for tweet ID, start date, end date, and tweet text. The tweet ID is a unique identifier for each tweet, while the start and end dates indicate the time range during which the tweet was collected. The tweet text column contains the content of the tweet.

## 4. METHODS

### 4.1 Sentiment Analysis of Twitter Data with VADER

To analyze the sentiment of the collected tweets, we utilized the VADER (Valence Aware Dictionary and sEntiment Reasoner) sentiment analysis tool. VADER is a lexicon-based sentiment analysis tool that uses a predefined set of rules to calculate sentiment scores for a given text. The lexicon consists of a list of words, each with an associated sentiment score. VADER also incorporates grammatical and syntactical rules to better understand the sentiment expressed in a text (Elbagir and Yang 2019; Hutto and Gilbert 2014).

VADER calculates sentiment scores by evaluating the sentiment of individual words in a text, taking into account the surrounding context. For example, it considers the impact of intensifiers (e.g., "very" or "extremely"), negations (e.g., "not" or "but"), and the sentiment of emojis and emoticons. VADER then combines these individual sentiment scores to compute the overall sentiment scores for the entire text (Hutto and Gilbert 2014).

Let's revisit a tweet example: "I love this vaccine! ". VADER would first assign sentiment scores to each word, known as valence scores in the tweet:

➔ "I": neutral (0.0)

➔ "love": positive (e.g., 0.7)

➔ "this": neutral (0.0)

➔ "vaccine": neutral (0.0)

➔ "!": neutral (0.0)

➔ "": neutral (0.0)

➔ "": positive (e.g., 0.3)

VADER then calculates the total sentiment scores for the tweet by combining the individual scores:

➔ Positive: 0.7 (love) + 0.3 (😊) = 1.0

➔ Negative: 0.0 (no negative words or emojis)

➔ Neutral: 0.0 (remaining words and emojis)

Finally, VADER computes the normalized sentiment scores and the compound score. The positive, negative, and neutral scores are normalized by dividing by the sum of the absolute values of all three scores, to ensure that they add up to 1. The compound score is calculated by summing the valence scores of each word in the lexicon, adjusted according to the rules, and then normalized to be between -1 and 1.

➔ Positive, negative, and neutral normalized scores:

➔ Positive: 1.0 / (1.0 + 0.0 + 0.0) = 1.0

➔ Negative: 0.0 / (1.0 + 0.0 + 0.0) = 0.0

➔ Neutral: 0.0 / (1.0 + 0.0 + 0.0) = 0.0

The compound score is calculated by summing the valence scores of each word in the lexicon, adjusted according to the rules, and then normalized to be between -1 and 1 using the following equation:

$$compound = \frac{\sum_{i}^{i} valence\ scores}{\sqrt{valence\ scores^2 + \alpha}}$$

Here, α is a normalization constant, typically set to 15 in the VADER implementation.

In the example tweet, "I love this vaccine!  ", the valence scores for each word and emoji are first calculated, and then the compound score is computed using the equation above. The compound score of 0.8 indicates a strong positive sentiment toward the vaccine.

By incorporating sentiment analysis into our dataset, we can further investigate the impact of news, companies, and CEOs on stock market performance. The sentiment scores provide valuable information about the public's perception of the topics and individuals being discussed, allowing us to examine how these perceptions may influence stock prices.

**4.2 Data Preprocessing of the stock market and sentiment scores, including correlation.**

To analyze the impact of news, companies, and CEOs on stock prices, we preprocessed the collected Twitter data and performed sentiment analysis as shown previously. The preprocessing steps included filtering the tweets to include only those related to weekdays when the stock market was open, as well as converting tweet data to New York Time. We then took the data for individual companies, CEOs, and news categories (COVID-19 and vaccines). We filtered the sentiment data based on the presence of company names, CEO names, or news categories. Stock data was then joined to these data based on the date and company name. The resulting data were further filtered to include only tweets from weekdays, between 9:30 AM and 3:30 PM, and from February 1, 2023, to March 19, 2023, to align with the stock market trading hours. Next, we calculated the average sentiment scores for each company, CEO, and news category. The average sentiment scores were then aggregated for each company, CEO, and news category at the same date and time. These averages are representing the sentiment scores for companies, CEOs, COVID-19 news, and vaccine news. For the competitor analysis, we calculated the average sentiment score of the other nine companies for each company at the same date and time. This allowed us to create data representing the sentiment scores for competitors, based on the simplified idea that for our 10 companies if we choose one the other nine are the competitors. This resulted in comprehensive data containing all the sentiment scores for each company, CEO, and news category, as well as the stock prices. This data served as the basis for our analysis of the impact of news, companies, and CEOs on stock market performance.

In addition to preprocessing the data, we also conducted a Pearson correlation analysis to explore the relationships between sentiment scores and stock market performance. This Pearson correlation analysis allowed us to better understand the linear associations between sentiment scores for different categories and stock market performance.

### 4.3 Modeling Stock Market Performance with time series model, ARIMA

The ARIMA (AutoRegressive Integrated Moving Average) model is a widely used time series forecasting technique that combines the concepts of autoregression (AR), differencing (I), and moving average (MA) to predict future values in a time series dataset (Kinney 1978; Newbold 1983). It is particularly useful for analyzing and forecasting univariate time series data, such as stock prices, which may exhibit a linear trend, seasonality, or a combination of both.

The ARIMA model consists of three components:

5. Autoregression (AR): This component captures the relationship between a data point and its previous values in the time series. In an AR model, the current value of a variable is assumed to be a linear function of its past values. The number of lagged terms used in the model is denoted by the parameter p.

6. Differencing (I): The differencing component is used to remove trends or seasonality in the data by subtracting the previous observation from the current one. The process is repeated d times, where d represents the order of differencing.

7.  Moving Average (MA): The moving average component models the error term as a linear combination of the error terms of the previous time steps. It captures the dependency between an observation and a residual error from a moving average model applied to lagged observations. The number of lagged error terms used in the model is denoted by the parameter q.

The ARIMA model is represented as ARIMA(p, d, q), where p, d, and q are the orders of the autoregression, differencing, and moving average components, respectively. To determine the optimal values for p, d, and q, the model selection process typically involves finding the set of parameters that minimize an information criterion, such as the Akaike Information Criterion (AIC) or the Bayesian Information Criterion (BIC). The Akaike Information Criterion (AIC) is a measure that evaluates the goodness of fit of a statistical model while penalizing the model's complexity. It helps to strike a balance between overfitting and underfitting, promoting model simplicity and accuracy. The model with the lowest AIC is considered the best among the candidate models. The Bayesian Information Criterion (BIC), on the other hand, is another measure used for model selection, which is similar to AIC but imposes a stronger penalty on model complexity. BIC is based on Bayesian principles and aims to select the model that is most likely to have generated the observed data. Like AIC, the model with the lowest BIC is considered the best choice among the competing models.Both AIC and BIC are commonly used in the model selection process to identify the most appropriate ARIMA model, as they help to find the optimal balance between fitting the data well and avoiding overfitting due to excessive complexity. By minimizing either the AIC or BIC, the chosen ARIMA model is expected to provide better forecasting accuracy and more reliable predictions (Kinney 1978).

In our study, we employed the ARIMA model to predict stock prices for each company in our dataset. To analyze the impact of news, companies, and CEOs on stock market performance, we used the comprehensive data generated from the preprocessing and sentiment analysis steps. First, we split the data into training and testing sets. The training set consisted of data from February 1, 2023, to March 7, 2023, accounting for 80% of our total data, while the testing set covered data from March 8, 2023, to March 19, 2023. The data from February 20th was removed, given that it was "Presidents' Day" in the USA, and the New York stock market did not open.

Due to the limited amount of data, we decided not to use a separate validation dataset. A validation dataset is usually used to fine-tune model parameters and prevent overfitting, which can occur when a model learns the training data too well and performs poorly on unseen data. However, with a low number of data points, splitting the data into three sets (training, validation, and testing) could lead to insufficient data for effective model training, which might compromise the model's performance.The disadvantage of not using a validation dataset is that it might be more challenging to identify and address overfitting or other potential issues that may arise during the model training process. Moreover, without a validation dataset, we may not have an unbiased evaluation of the model's performance before testing it on the test dataset. For future analyses, we plan to incorporate a validation dataset, especially when working with larger datasets. This will enable us to fine-tune our model parameters more effectively and provide a more accurate assessment of the model's performance, leading to more reliable predictions and insights into the impact of news, companies, and CEOs on stock market performance.

Next, we fit an ARIMA model for each company using the auto.arima function from the forecast package in R. The auto.arima function automatically selects the best ARIMA model based on the Akaike Information Criterion (AIC) and allows for non-seasonal time series modeling. The AIC is a statistical measure used to compare different models, taking into account both the goodness of fit and the complexity of the model. The auto.arima function examines a range of possible ARIMA models and selects the one with the lowest AIC value, indicating the best balance between fitting the data and avoiding overfitting. By using auto.arima, we can automatically select the best-fitting ARIMA model for each company's time series data without the need for manual inspection or trial-and-error. This makes the modeling process more efficient and reliable, as it minimizes human error and subjectivity in model selection. After fitting the ARIMA models, we made predictions for each company on the testing data set.

Finally, we evaluated the performance of our ARIMA models by calculating the Mean Absolute Percentage Error (MAPE) for each company. The MAPE is a measure of the accuracy of the predictions compared to the actual values, expressed as a percentage. It is calculated by taking the absolute difference between the predicted and actual values, dividing this difference by the actual value, and then averaging these percentages across all data points.

The formula for MAPE is as follows:

$$M = \frac{1}{n} \sum_{t=1}^{n} \left| \frac{A_t - F_t}{A_t} \right|$$

$M$ = mean absolute percentage error
$n$  = number of times the summation iteration happens
$A_t$ = actual value
$F_t$ = forecast value

The MAPE is a popular metric for evaluating the performance of time series models, as it is easily interpretable and provides a clear indication of the average error rate in terms of percentage. Lower MAPE values indicate better model performance, as they imply smaller deviations between the predicted and actual values.

We calculated the MAPE for each company and then computed the overall mean MAPE to assess the performance of our time series models in predicting stock prices. This evaluation step helps us understand the effectiveness of our models and provides insights for potential improvements or alternative modeling approaches in future analyses.

**4.4 Incorporating Sentiments as Covariates in Stock Market Prediction**

In this section, we extended our ARIMA models by incorporating company-specific, COVID-19-related, vaccine-related, CEOs, competitors' sentiments as covariates. The goal was to assess the influence of these factors on stock market performance.

We split the data into training and testing sets as before, and fit separate ARIMA models for each company, this time including the relevant sentiment covariates. The auto.arima function was utilized to fit these models, as in the previous analysis. Subsequently, we made predictions on the testing set for each company To evaluate the performance of our models with sentiment covariates, we calculated the Mean Absolute Percentage Error (MAPE) for each company, as well as the overall mean MAPE. This enabled us to assess the predictive accuracy of our time series models when incorporating company-specific, COVID-19, vaccine-related, CEOs, competitors' sentiments as covariates. By comparing the MAPE values for each sentiment factor, we can determine the extent to which these sentiments influence stock market performance. Additionally, we fit a second set of models that consider all the factors and another that only consider company sentiment and vaccine sentiment as covariates, because they were the factors that performed the best in the previous models. We then compare the performance of these two sets of models by comparing their respective MAPE values.

### 4.5 Modeling Stock Market Performance with Time Series Model, VAR

The Vector Autoregression (VAR) model is a multivariate time series model that captures the linear interdependencies among multiple time series variables (Stock and Watson 2001). It is an extension of the univariate autoregressive (AR) model to multiple variables, allowing for the simultaneous modeling and prediction of several related time series. VAR models are particularly useful in the analysis of economic, financial, and other systems where the variables are influenced by each other.

A VAR model consists of a system of equations, with each equation representing a time series variable. Each variable in the system is expressed as a linear combination of its own lagged values and the lagged values of the other variables. The number of lagged terms used in the model is denoted by the parameter p, representing the order of the VAR model (VAR(p)).

The general form of a VAR(p) model for k variables is as follows:

$$Y(t) = c + A1 * Y(t-1) + A2 * Y(t-2) + \dots + Ap * Y(t-p) + \varepsilon(t)$$

Here, Y(t) is a k-dimensional vector of the time series variables at time t, c is a k-dimensional vector of constants, Ai (i = 1, 2, ..., p) are k × k matrices of coefficients, and ε(t) is a k-dimensional vector of error terms assumed to be normally and independently distributed with zero mean and constant variance(Stock and Watson 2001).

VAR models are widely used in economics, finance, and other fields for various purposes, including understanding the relationships among variables, impulse response analysis, and Granger causality testing. They provide a flexible and powerful tool for modeling and forecasting multivariate time series data, particularly when the variables exhibit interdependencies and feedback effects. Impulse response analysis is a technique that examines the response of a variable to a shock or impulse in another variable within a multivariate system. In the context of VAR models, it investigates how a one-time shock to one of the variables affects the other variables in the system over time. The impulse response functions trace the effects of a unit shock to one of the variables on the current and future values of all the variables in the system. This analysis is particularly useful for understanding the dynamic relationships and feedback effects among the variables in a system, providing insights into the speed, magnitude, and duration of the responses to shocks. Granger causality testing, on the other hand, is a statistical hypothesis test that examines whether past values of one variable can help predict the future values of another variable, beyond the information contained in the past values of the second variable itself. In other words, it tests if one variable "Granger-causes" another. This method is valuable for identifying causal relationships or predictive power between the variables in a VAR model, helping to establish the direction of causality and improve model interpretation. Together, impulse response analysis and Granger causality testing provide valuable insights into the complex dynamics of multivariate time series data, enabling more accurate and informative modeling, forecasting, and decision-making in various fields, such as economics and finance(Stock and Watson 2001).

In our study, we employed the VAR model to predict stock prices for each company in our dataset, while simultaneously accounting for the influence of other variables, such as company-specific, COVID-19-related, vaccine-related, CEOs, and competitor sentiments. To analyze the impact of these factors on stock market performance, we used the comprehensive data generated from the preprocessing and sentiment analysis steps. To implement the VAR model, we used the vars package from R, which provides various tools for fitting and analyzing VAR models. First, we split the data into training and testing sets. As before, the training set consisted of data from February 1, 2023, to March 7, 2023, accounting for 80% of our total data, while the testing set covered data from March 8, 2023, to March 19, 2023. As in the ARIMA analysis, we excluded the data from February 20th due to the US Presidents' Day holiday, and we decided not to use a separate validation dataset due to the limited amount of data. Next, we fit a VAR model for each company using the vars function from the vars package in R. The vars function enables us to fit VAR models of different orders (p) and determine the optimal order using an information criterion, such as the Akaike Information Criterion (AIC) or the Bayesian Information Criterion (BIC). This allows us to select the best-fitting VAR model for each company's time series data without the need for manual inspection or trial-and-error, improving the efficiency and reliability of the modeling process. After fitting the VAR models, we made predictions for each company on the testing data set. We then evaluated the performance of our VAR models by calculating the Mean Absolute Percentage Error (MAPE) for each company, as well as the overall mean MAPE. This evaluation step helps us assess the effectiveness of our models and provides insights for potential improvements or alternative modeling approaches in future analyses. Additionally, we fit another model that considered all the factors at once.

## 5. RESULTS AND DISCUSSION

### 5.1 Volatility of Stock Prices in After-Hours Trading

Based on the analysis of historical stock market data for the ten studied biotech companies, we found that stock prices were more volatile during after-hours trading (Figure 1). After-hours trading refers to the time outside the regular trading hours of the New York Stock Exchange (NYSE), which are from 9:30 AM to 4:00 PM. In our study, we collected data until 3:30 PM. It is important to note that after-hours trading also includes weekends, as stock prices may be influenced by events and news occurring outside regular trading hours.

This increased volatility during after-hours trading was expected, as the lower trading volume and reduced liquidity during these times often result in larger price fluctuations. Additionally, significant news or events that take place outside regular trading hours can have a considerable impact on stock prices, contributing to the observed increased volatility.

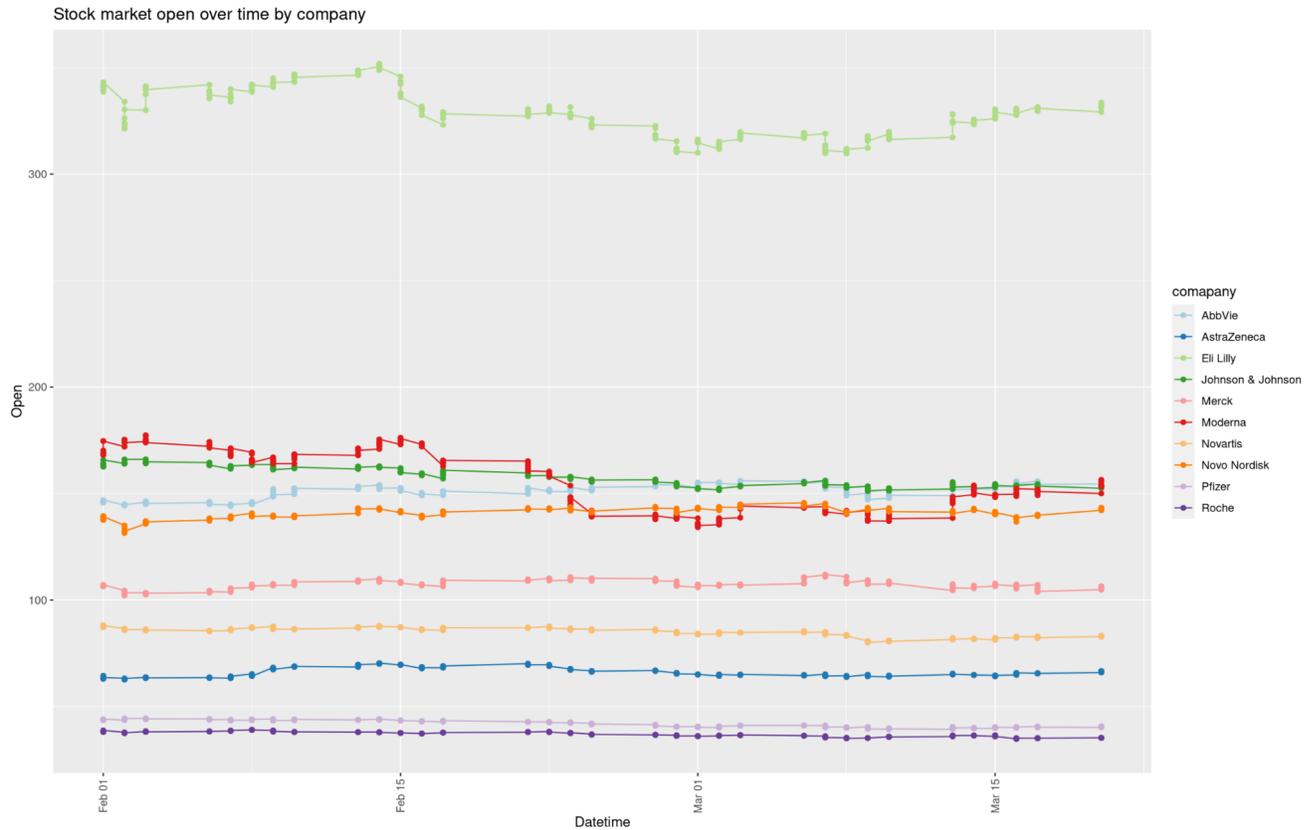

Figure 1: Daily stock market opening prices from February 1st to March 19th, and hourly opening prices from 9:30 AM to 3:30 PM for the 10 companies included in the study.

### 5.2 Analysis of Collected Twitter Data on News, Companies, and CEOs

Through our collection of tweets related to COVID-19, vaccines, the ten biotech companies, and their respective CEOs, we aimed to validate our hypothesis that news, companies, and CEOs affect the stock market. Analyzing the daily number of collected tweets related to the companies in our study, spanning from February 1st to March 19th, we observed the volume of tweets mentioning each company, CEOs, COVID-19, and vaccines during the time period.

Our results indicate that in terms of company mentions, Roche, Pfizer, and Moderna consistently received the highest volume of tweets per day, with an average of around 700 tweets daily throughout the entire period, which is the maximum possible we can collect, given the maxim tweets able to call by the Twitter API is 100 per hour and we are analyzing 7 hours (Figure 2). In the case of CEOs, Moderna and Pfizer's CEOs attracted the most attention, with some days reaching a maximum of 700 tweets, in contrast to the rest of the CEOs, whose tweet counts never exceeded 160 tweets per day (Figure 3). Regarding news mentions, both vaccine and COVID-19-related tweets had high counts (Figure 4). On one day, the minimum count for such tweets was 6642, while on other days, the range was between 660 and 680 tweets (Figure 4). For the rest of the days, the tweet counts exceeded 680 tweets per day (Figure 4).

These results provide insight into the public's engagement and discussions surrounding the biotech companies and their CEOs, as well as COVID-19 and vaccine-related news. The higher volume of tweets for certain companies and CEOs may suggest a more significant impact on their stock market performance, as they capture more public attention and sentiment.

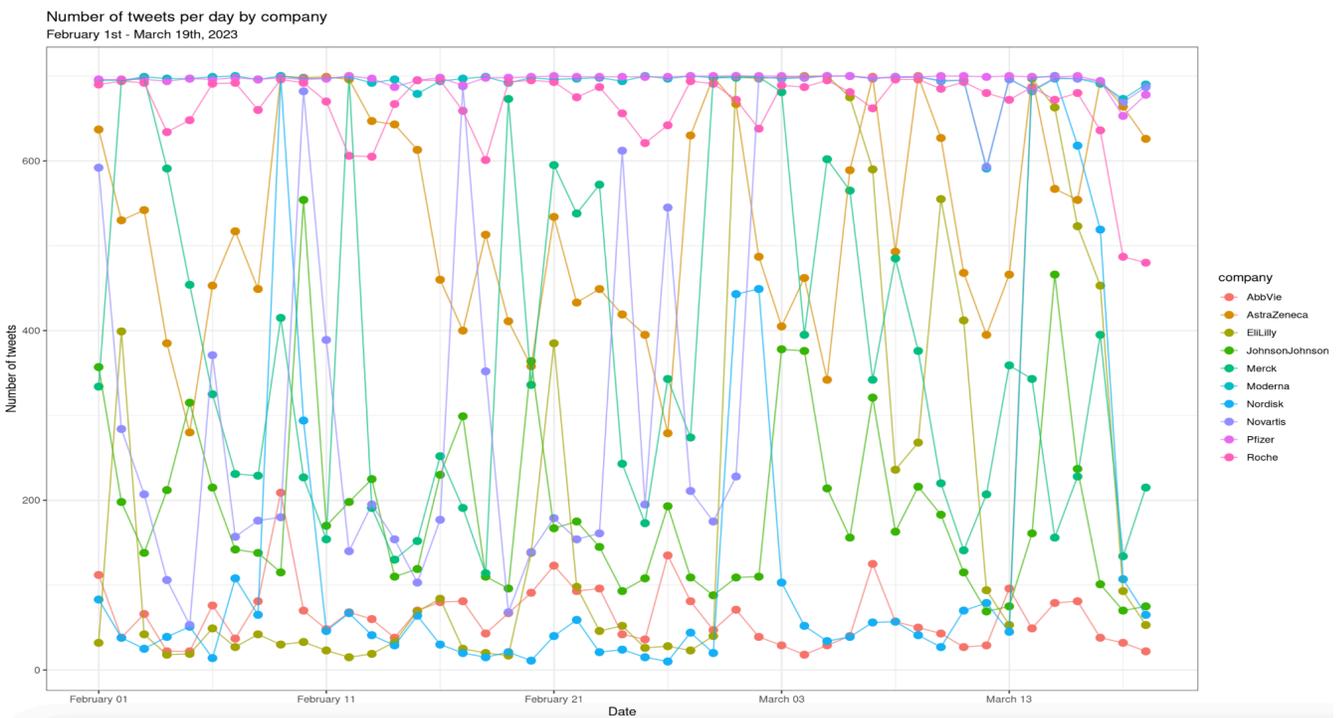

Figure 2: Daily number of collected tweets related to the companies in our study, spanning from February 1st to March 19th. This figure highlights the volume of tweets mentioning each company during the time period.

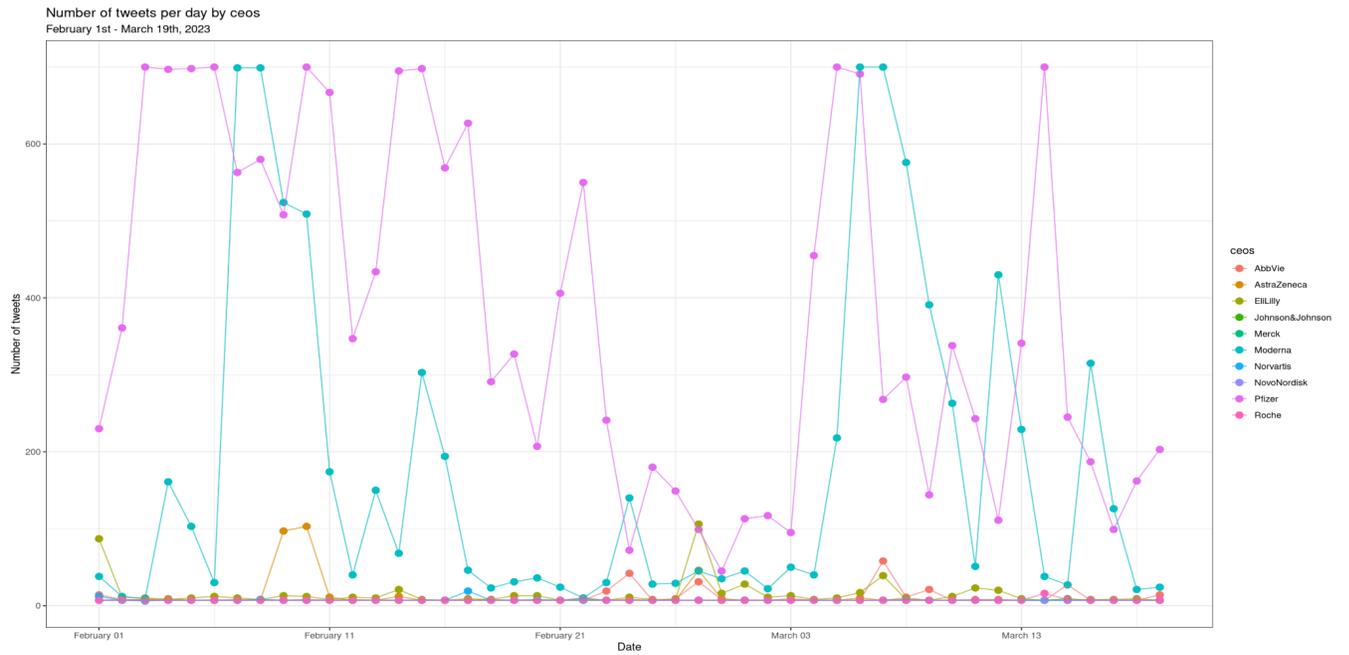

Figure 3: Daily number of collected tweets related to the companies in our study, spanning from February 1st to March 19th. This figure highlights the volume of tweets mentioning each CEOs during the time period.

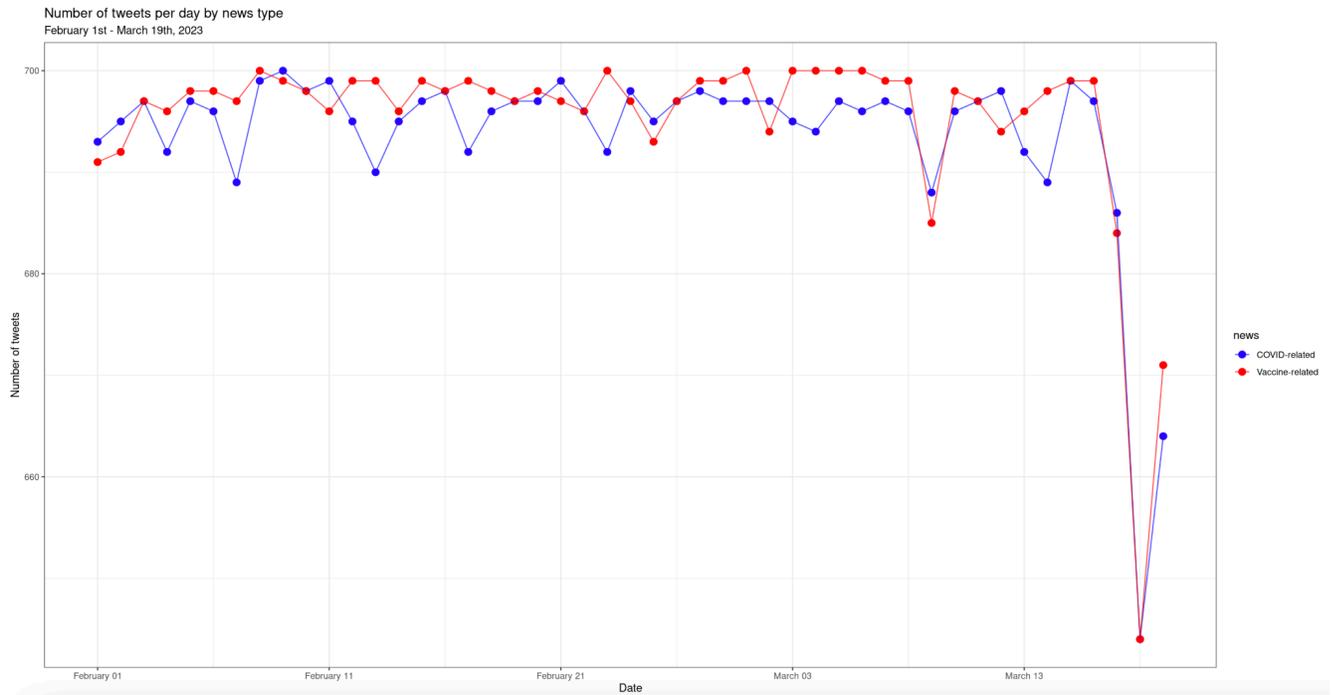

Figure 4: Daily number of collected tweets related to the companies in our study, spanning from February 1st to March 19th. This figure highlights the volume of tweets mentioning news related to COVID and vaccine during the time period.

### 5.3 Exploring Public Sentiment Patterns with VADER Sentiment Analysis

Our study utilized the VADER sentiment analysis tool to analyze the sentiment of the collected tweets related to COVID-19, vaccines, the ten biotech companies, and their respective CEOs. The sentiment scores provide valuable information about the public's perception of the topics and individuals being discussed, allowing us to examine how these perceptions may influence stock prices.

The negative, neutral, and positive sentiment scores obtained from the sentiment analysis of tweets across all categories, including companies, CEOs, and news, revealed a distinct pattern. The majority of negative and positive sentiment scores were concentrated in the range of 0 to 0.25, while neutral sentiment scores predominantly fell between 0.75 and 1(Figures 5, S1, and S2). This pattern suggests that the tweets in our dataset tended to express either mild positive or negative emotions or predominantly neutral sentiments. It may indicate that the public discourse surrounding these biotech companies, their CEOs, and related news events was not highly polarized or emotionally charged. However, some specific companies and CEOs had a higher density of tweets with more pronounced positive sentiment scores, which might reflect favorable public opinions toward them.

Upon examining the overall sentiment scores (that goes from -1 to 1) for tweets, which were derived as a compound value from the positive, neutral, and negative sentiment scores, we identified several noteworthy patterns. In tweets related to companies (Figure 6), the majority of tweets had sentiment scores around 0, suggesting a neutral sentiment. However, AbbVie, Eli Lilly, and Novo Nordisk emerged as exceptions, displaying a higher density of tweets with sentiment scores between 0.1 and 1, indicating a more favorable perception of these companies. Similarly, in the case of tweets related to CEOs, most tweet sentiment scores were at 0, denoting neutral opinions. Interestingly, Eli Lilly and Pfizer stood out with a higher density of tweets bearing sentiment scores between 0.1 and 1, suggesting a more positive public sentiment towards their respective CEOs. When considering tweets related to news, a comparison between COVID-19 and vaccine-related tweets showed a distinct contrast. Vaccine tweets had a higher density of sentiment scores between 0.1 to 1 and -0.1 to -1, signifying a more polarized distribution of opinions on this subject compared to COVID-19 tweets. These findings imply that public sentiment towards biotech companies, their CEOs, and news events can vary significantly. While some companies and CEOs receive more favorable public opinions, certain topics, such as vaccines, tend to provoke more polarized reactions.

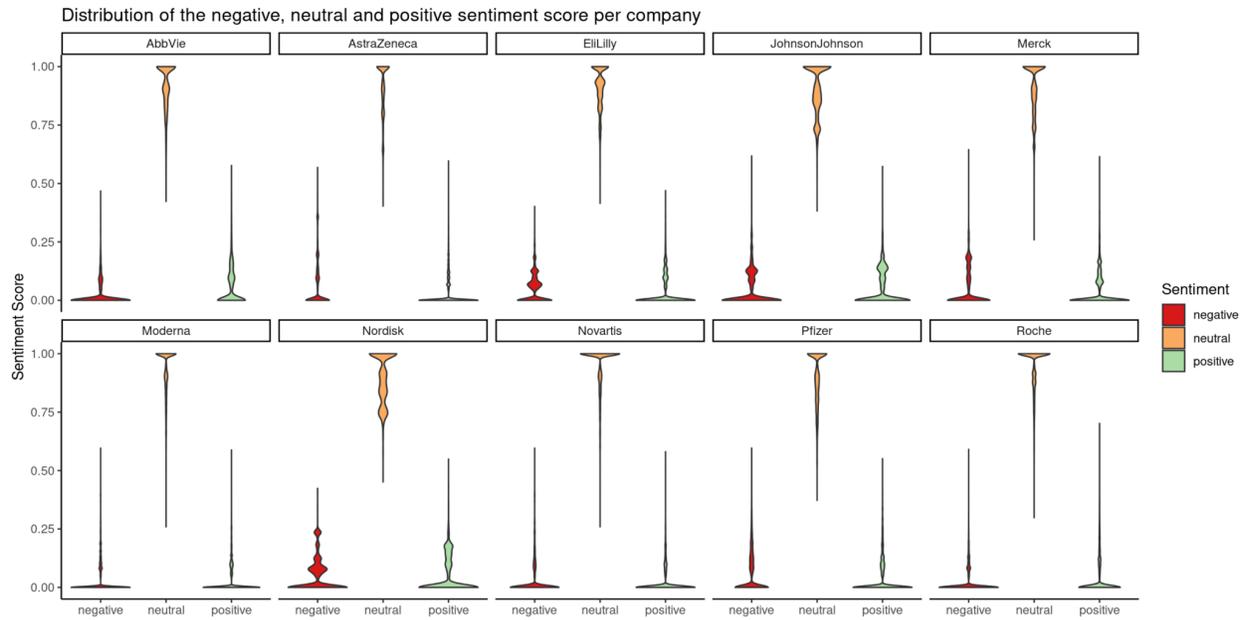

Figure 5: A distribution showcasing the negative, neutral, and positive sentiment scores obtained from the sentiment analysis of tweets related to the companies in our study. This figure provides a visual understanding of the sentiment breakdown for each company.

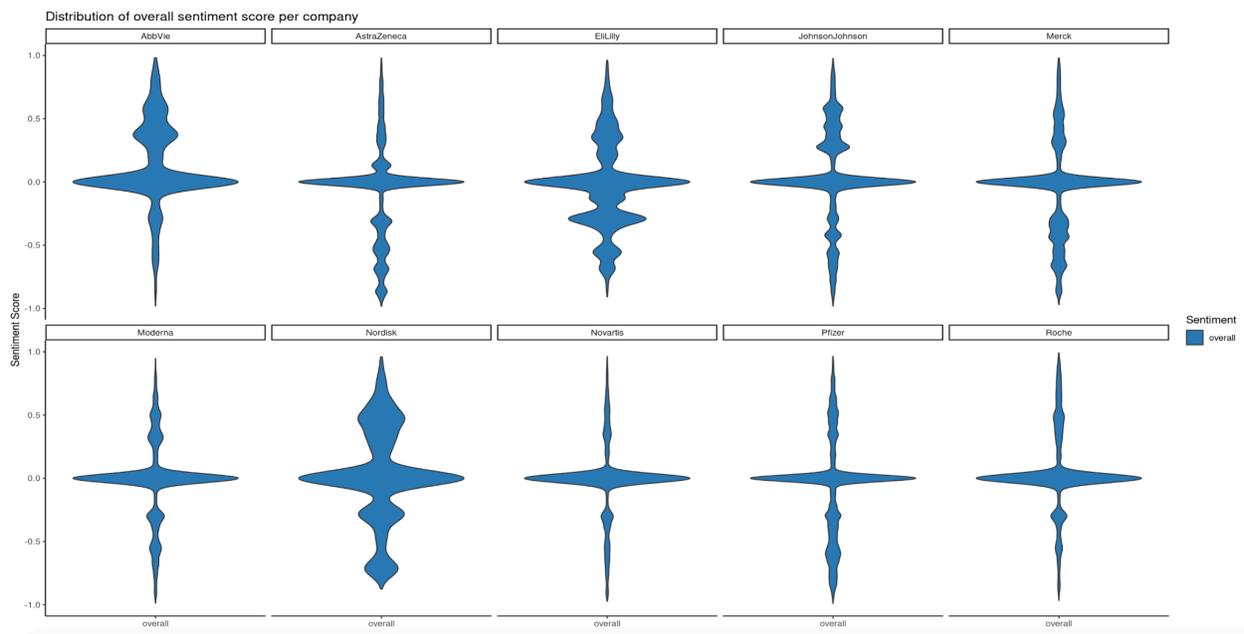

Figure 6: A distribution of the overall sentiment scores for tweets related to companies. These scores are calculated as a compound value derived from the positive, neutral, and negative sentiment scores. This figure offers insights into the general sentiment trends for each company.

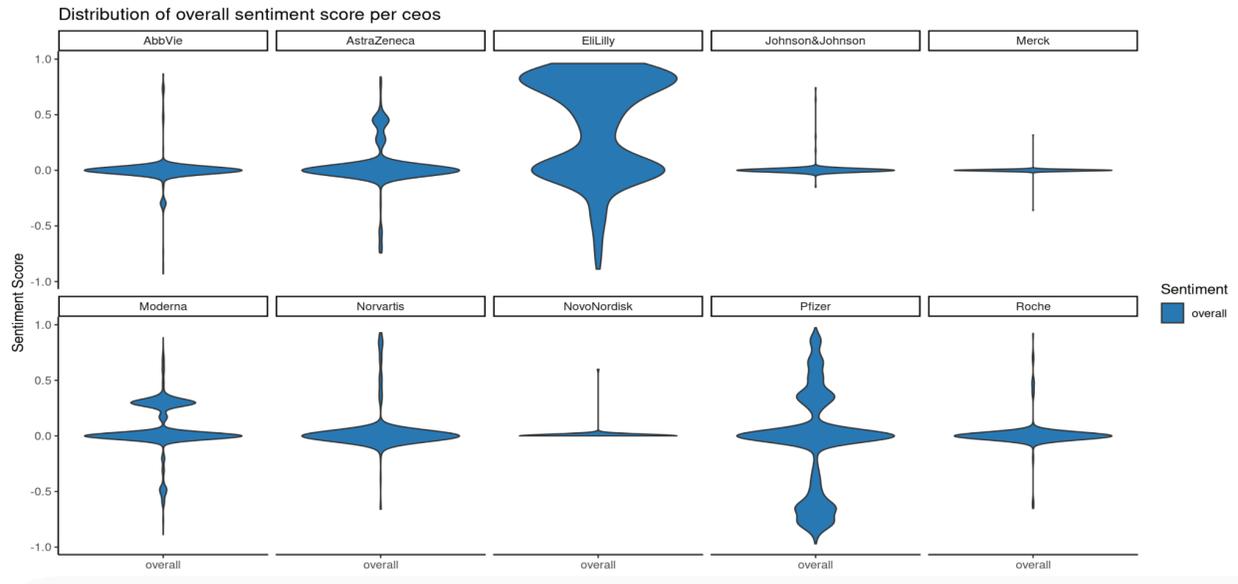

Figure 7: A distribution of the overall sentiment scores for tweets related to companies. These scores are calculated as a compound value derived from the positive, neutral, and negative sentiment scores. This figure offers insights into the general sentiment trends for each CEOs.

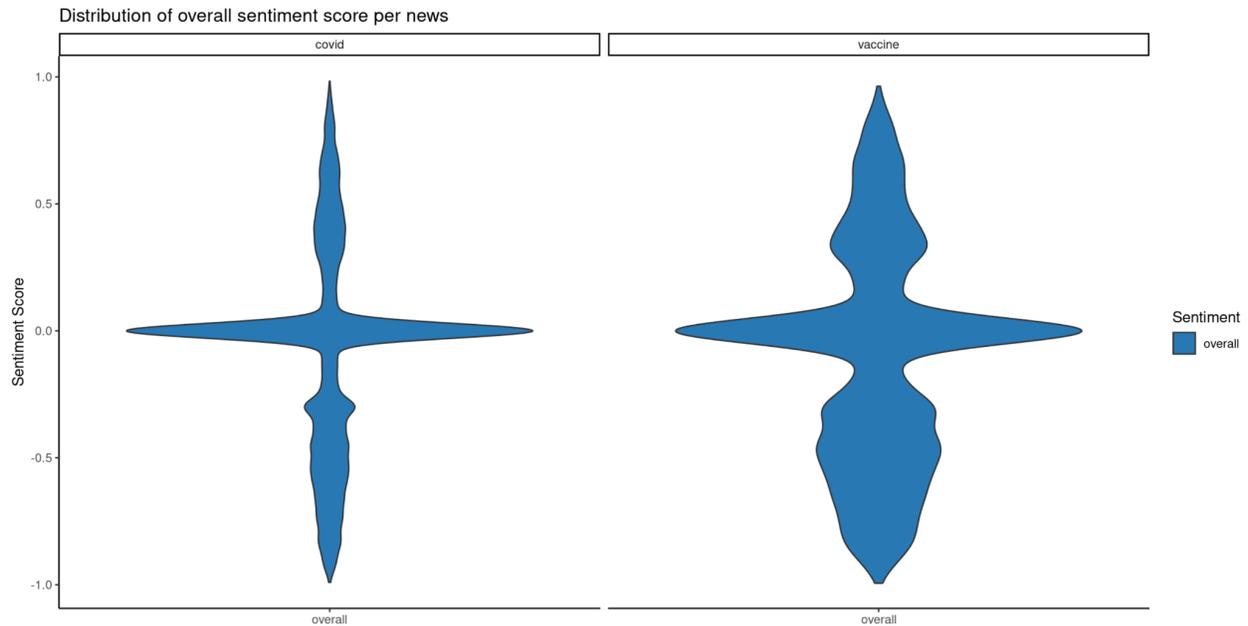

Figure 8: A distribution of the overall sentiment scores for tweets related to companies. These scores are calculated as a compound value derived from the positive, neutral, and negative sentiment scores. This figure offers insights into the general sentiment trends for each news which include COVID and vaccines.

### 5.4 Relationship between sentiment scores

The sentiment scores of the company, COVID-19, vaccine, CEOs, and competitors exhibit a normal distribution (Figure 9). It is important to note that normal distribution is not a strict requirement for time series forecasting methods like ARIMA. However, having normally distributed data can be beneficial, as it simplifies the modeling process and ensures that the model assumptions are met. In the case of ARIMA, it assumes that the residuals (errors) are normally distributed. If the data is not normally distributed, transformations can be applied to improve the distribution and satisfy the model's assumptions. Nevertheless, even with non-normally distributed data, ARIMA can still provide useful forecasts, but the model's performance might be affected.

The correlation analysis (Figue 10) provided insightful results regarding the relationships between sentiment scores of different categories and stock market performance. The most significant result was found between company sentiment scores (companyS) and competitor sentiment scores (competitorsS) for Merck (r = 0.313), Moderna (r = 0.238), and Novartis (r = 0.204). This implies that for these companies, the sentiment scores of their competitors have a notable positive correlation with their own sentiment scores. This finding can be speculated by the fact that these companies operate likely to produce similar products and may experience similar events or factors affecting their businesses. As a result, the market sentiment toward one company might influence the sentiment toward others within the same industry. This correlation highlights the importance of analyzing the sentiment scores of competitors in addition to the target companies' sentiment scores when trying to understand the factors that could affect stock prices. It is also noteworthy that the categories do not show high correlations, which indicates that there is no redundancy in the information provided by each category. This is important because it suggests that each category (company, CEO, COVID-19 news, vaccine news, and competitors) contributes unique information to the overall sentiment analysis. This lack of redundancy strengthens our confidence in the sentiment analysis results and highlights the need to consider a variety of factors and perspectives when trying to understand the complex relationship between news, companies, CEOs, and stock market performance.

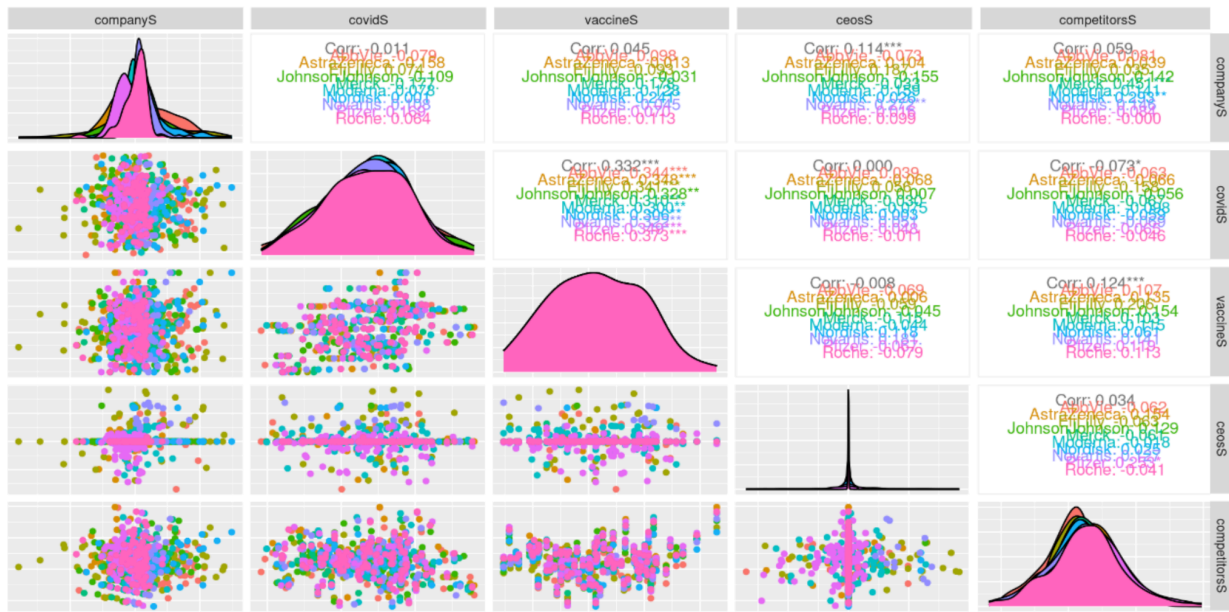

Figure 9: Distribution and relationships between overall sentiment scores for company, CEO, news (COVID, vaccine), and competitor sentiments. These scores are obtained from sentiment analysis of tweets and are grouped by companies. Correlation values are small and difficult to read in this figure but can be found in a larger size in Figure 10.

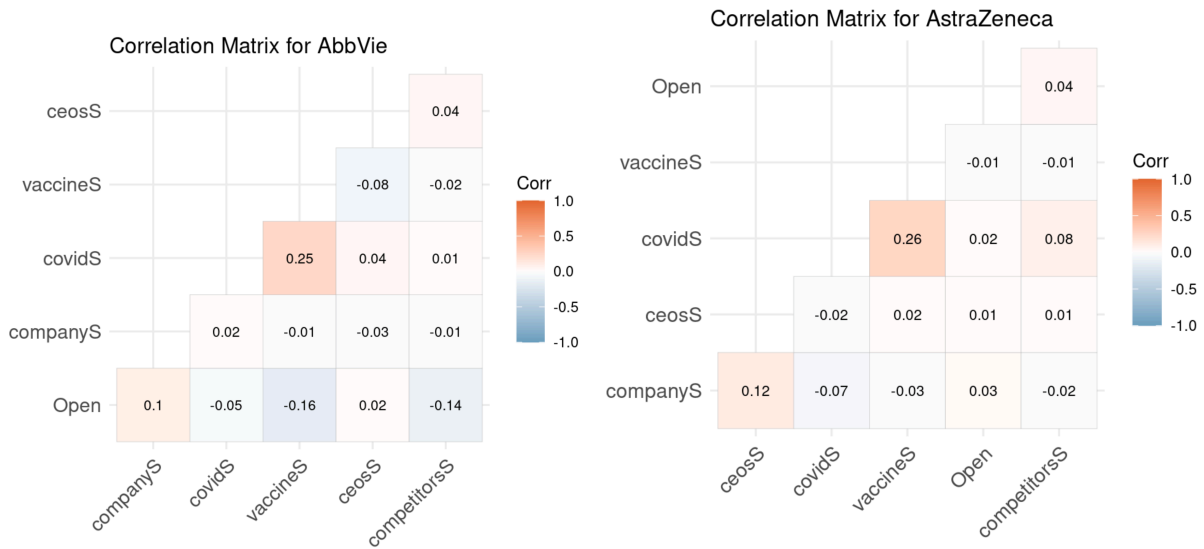

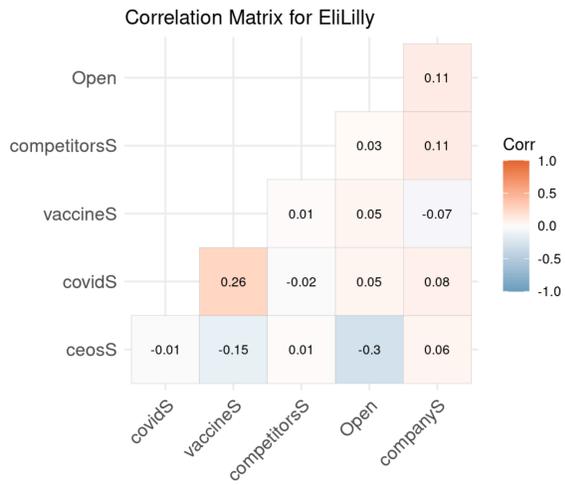

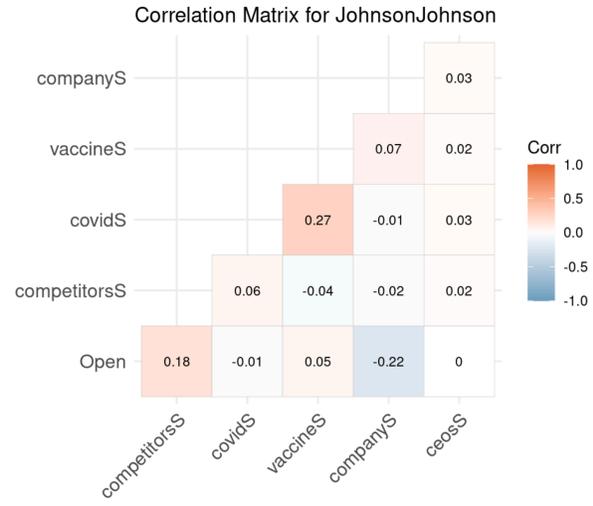

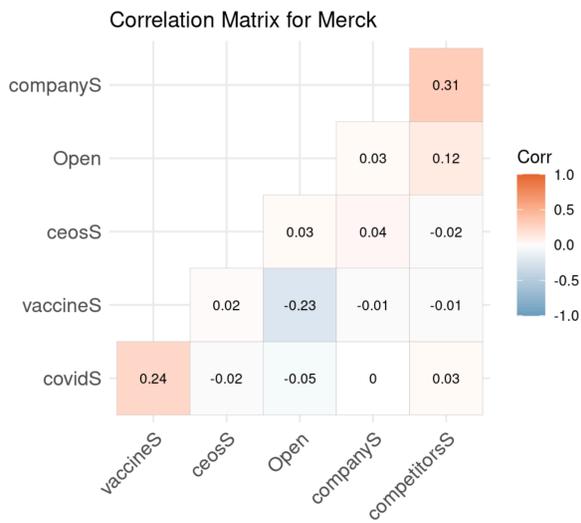

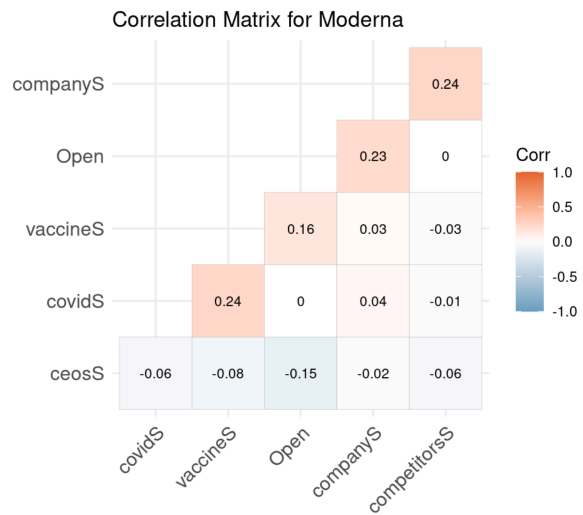

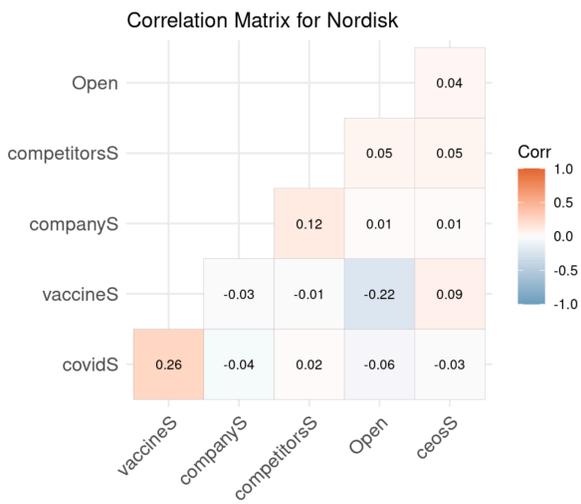

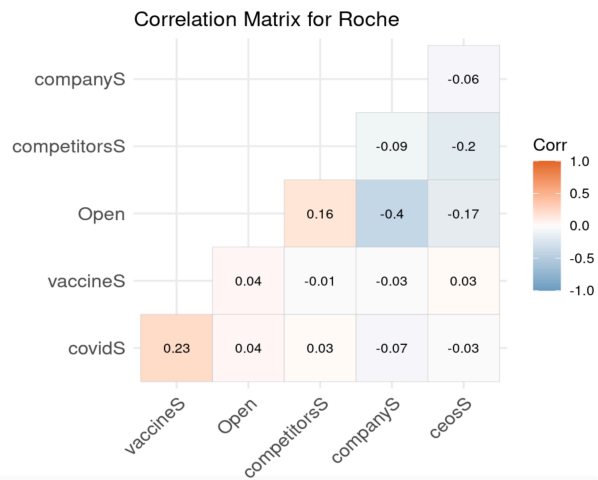

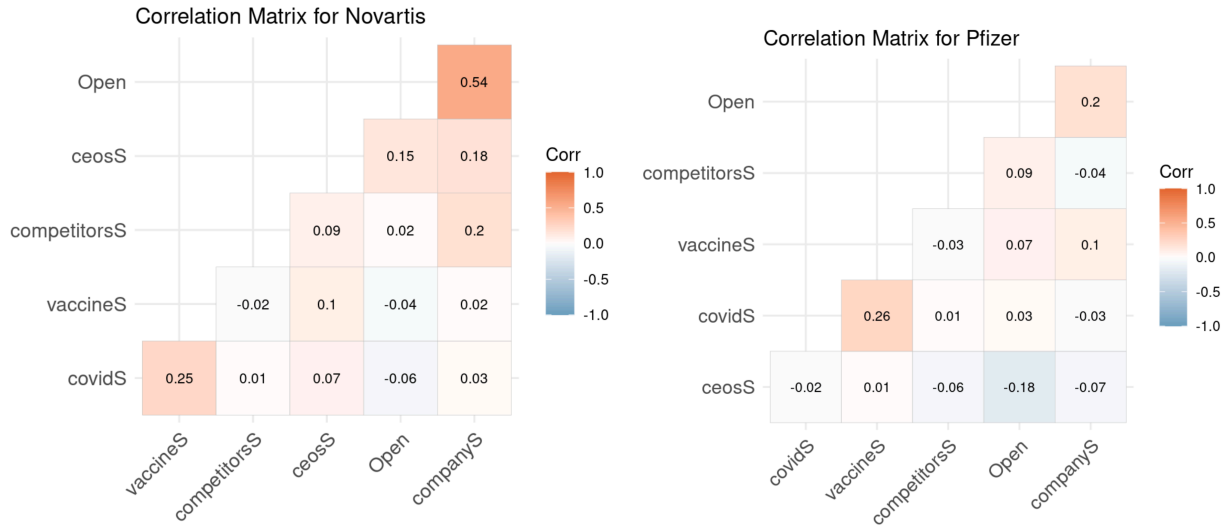

Figure 10: Correlation matrices displaying the relationships between the opening stock market values and the overall sentiment scores for the various categories: company, CEOs, COVID, vaccine, competitors, and including open prices of the stock market, for each pharmaceutical company.

### 5.5 Forecasting Predictions using ARIMA Model

In this section, we first evaluate the performance of ARIMA models using only historical records (Figure 11) for predicting stock market performance. Subsequently, we assess the effectiveness of incorporating sentiment covariates into the ARIMA models to determine if they lead to improved predictions (Figure 12-18). The Mean Absolute Percentage Error (MAPE) is employed as the key metric to compare the accuracy of our forecasts, analyzing the performance of ARIMA models with different covariates for each company. Our analysis aims to identify which covariates contribute to better predictions. By examining the impact of these sentiment covariates alongside historical records, we provide valuable insights into the complex relationship between sentiment and stock market performance. From the analysis, it can be observed that incorporating certain covariates, such as companies and vaccines, can help to achieve better predictions compared to using only historical records.

Taking a closer look at individual company performances, the results of our analysis indicate that the Vaccine covariate played a significant role in improving the stock market performance predictions for AstraZeneca and AbbVie (Figure 12). For AstraZeneca, the best prediction was achieved when using the Vaccine covariate, with a Mean Absolute Percentage Error (MAPE) of 1.2558, which was considerably lower than the historical record MAPE of 1.3205 (Table 1). Likewise, for AbbVie, the Vaccine covariate outperformed other factors, with a MAPE of 1.7093 compared to the historical record MAPE of 1.7655 (Table 1). This improved prediction accuracy may be attributed to the polarized opinions on vaccines observed in our sentiment analysis. Vaccine tweets exhibited a higher density of sentiment scores ranging from 0.1 to 1 and -0.1 to -1, indicating a more polarized distribution of opinions on this subject compared to the more neutral sentiment observed, such as example, COVID-19 tweets (Figure 8). The polarization of opinions on vaccines may lead to more significant fluctuations in investor sentiment, subsequently affecting stock market performance. This heightened sensitivity to vaccine-related news and opinions may explain the superior performance of the Vaccine covariate in predicting stock market outcomes for AstraZeneca and AbbVie, as it captures the strong influence of these polarized views on investor decision-making and stock prices. In light of these findings, it is essential to consider the impact of polarized opinions, particularly in the context of vaccine-related news, when modeling and predicting stock market performance. Recognizing and accounting for these polarized views could provide a more accurate understanding of the relationship between news sentiment and stock market behavior, ultimately leading to better-informed investment decisions and strategies. For other companies, such as Eli Lilly, Johnson, and Merck, a combination of all covariates seems to yield better predictions than any single category (Table 1).

In summary, our analysis demonstrates that incorporating news sentiment covariates into the ARIMA models leads to improved predictions of stock market performance. The results highlight the importance of considering various factors, such as company news, CEO news, vaccine news, COVID news, and competitor news, when modeling and predicting stock prices. Although the Vaccine covariate generally performs better, it is crucial to consider other covariates and their combinations to achieve the most accurate predictions for each specific company.

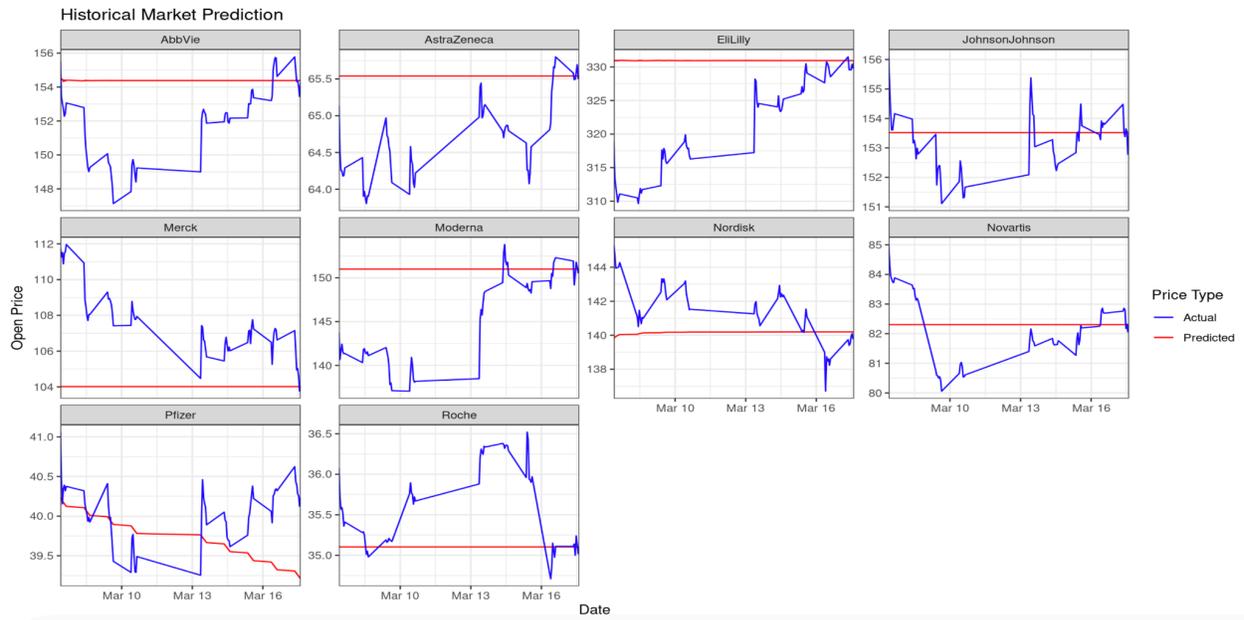

Figure 11: Comparison of actual and predicted stock market opening prices from March 8th to March 17th for individual companies, as forecasted by the ARIMA model using historical market opening data.

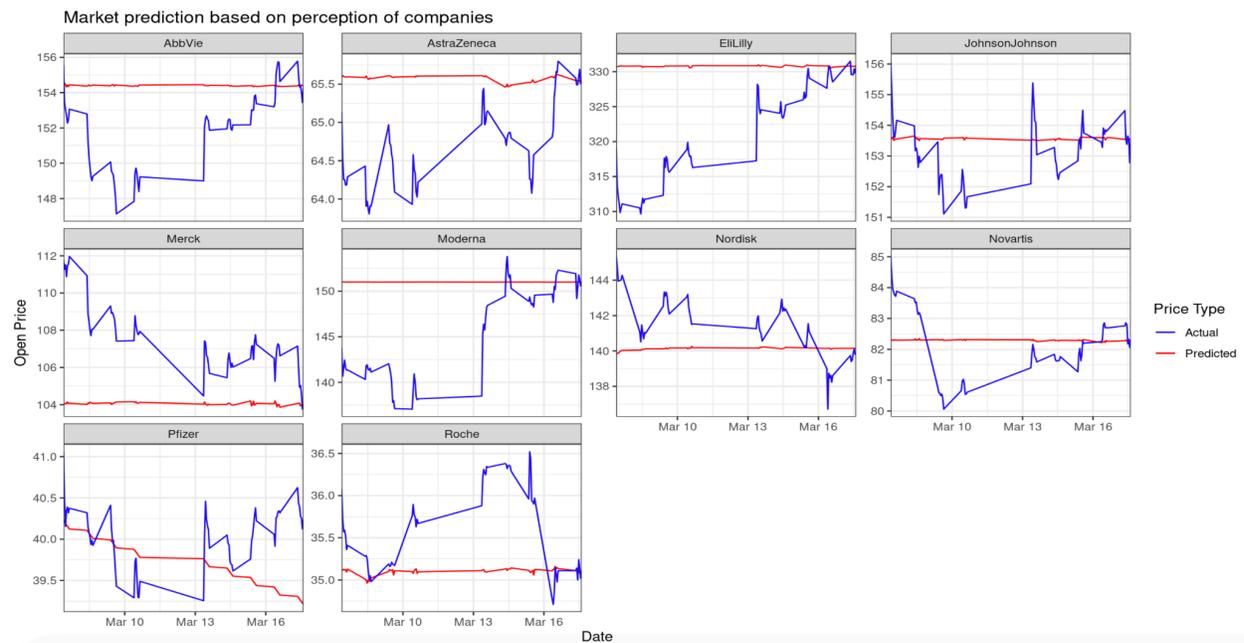

Figure 12: A comparison of the actual and predicted stock market opening prices for each company from March 8th to March 17th, as forecasted by the ARIMA model incorporating historical market opening data and using sentiment analysis of tweets related to the companies as covariates. This figure visually demonstrates the accuracy and performance of the ARIMA model in predicting stock market opening prices while considering the influence of public sentiment as covariates in the model.

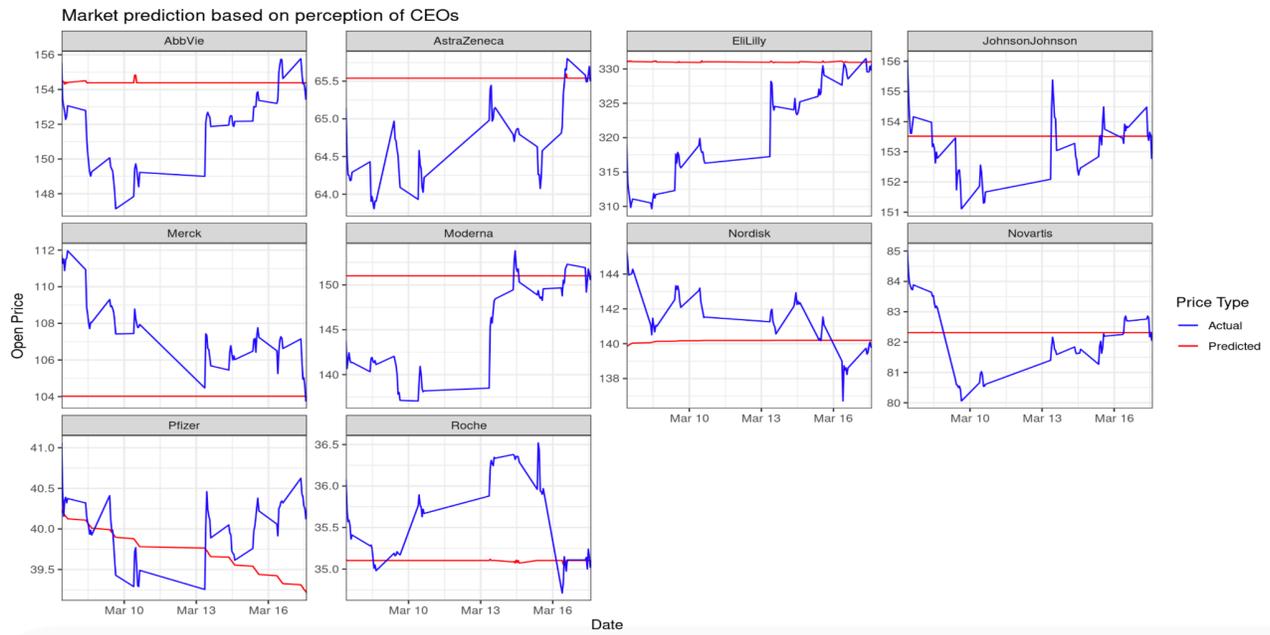

Figure 13: A comparison of the actual and predicted stock market opening prices for each company from March 8th to March 17th, as forecasted by the ARIMA model incorporating historical market opening data and using sentiment analysis of tweets related to the CEOs as covariates.

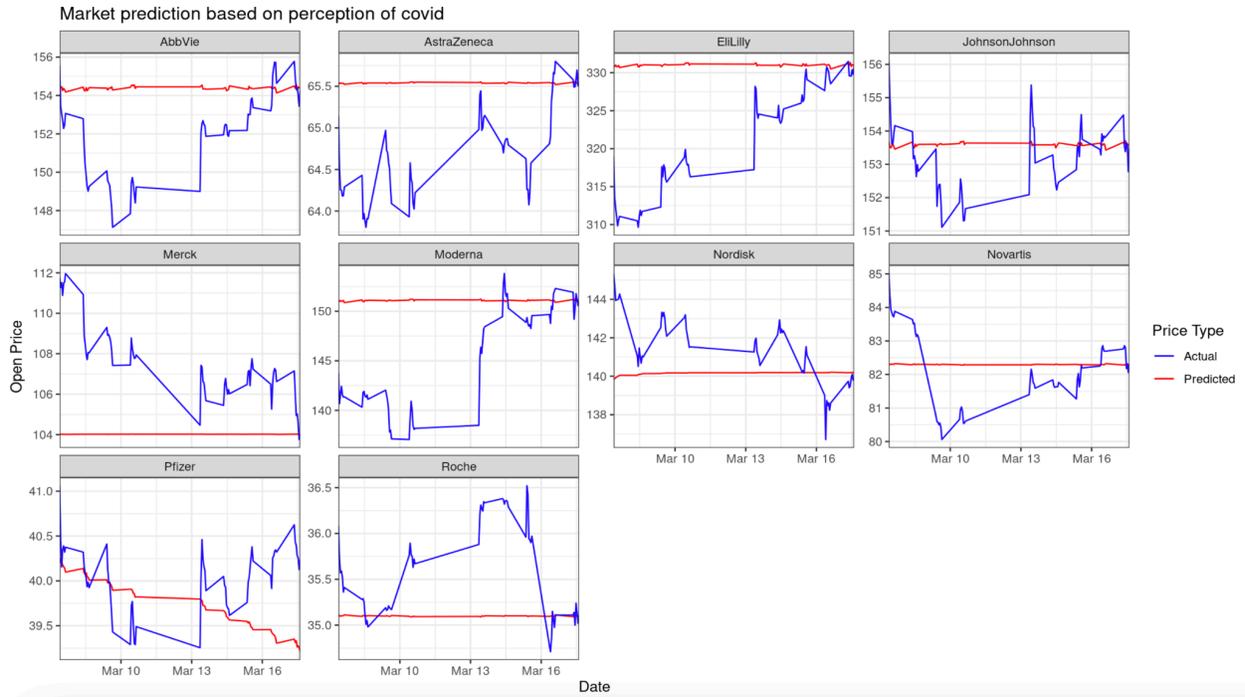

Figure 14: A comparison of the actual and predicted stock market opening prices for each company from March 8th to March 17th, as forecasted by the ARIMA model incorporating historical market opening data and using sentiment analysis of tweets related to the COVID as covariates.

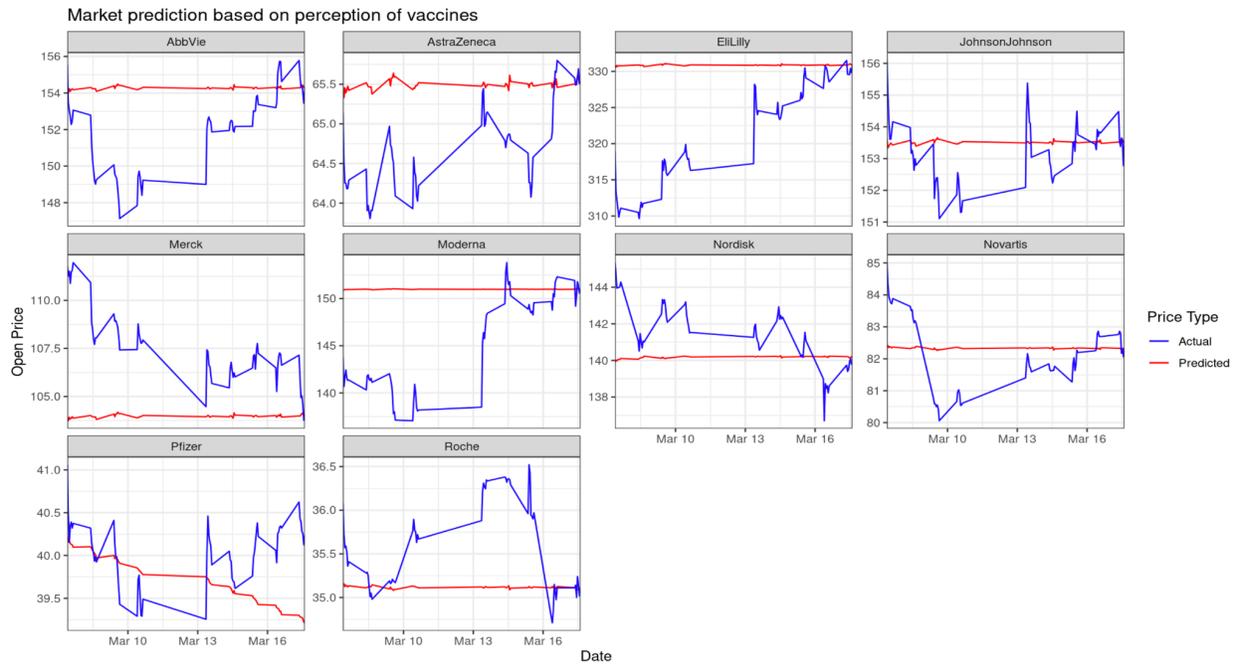

Figure 15: A comparison of the actual and predicted stock market opening prices for each company from March 8th to March 17th, as forecasted by the ARIMA model incorporating historical market opening data and using sentiment analysis of tweets related to the vaccines as covariates.

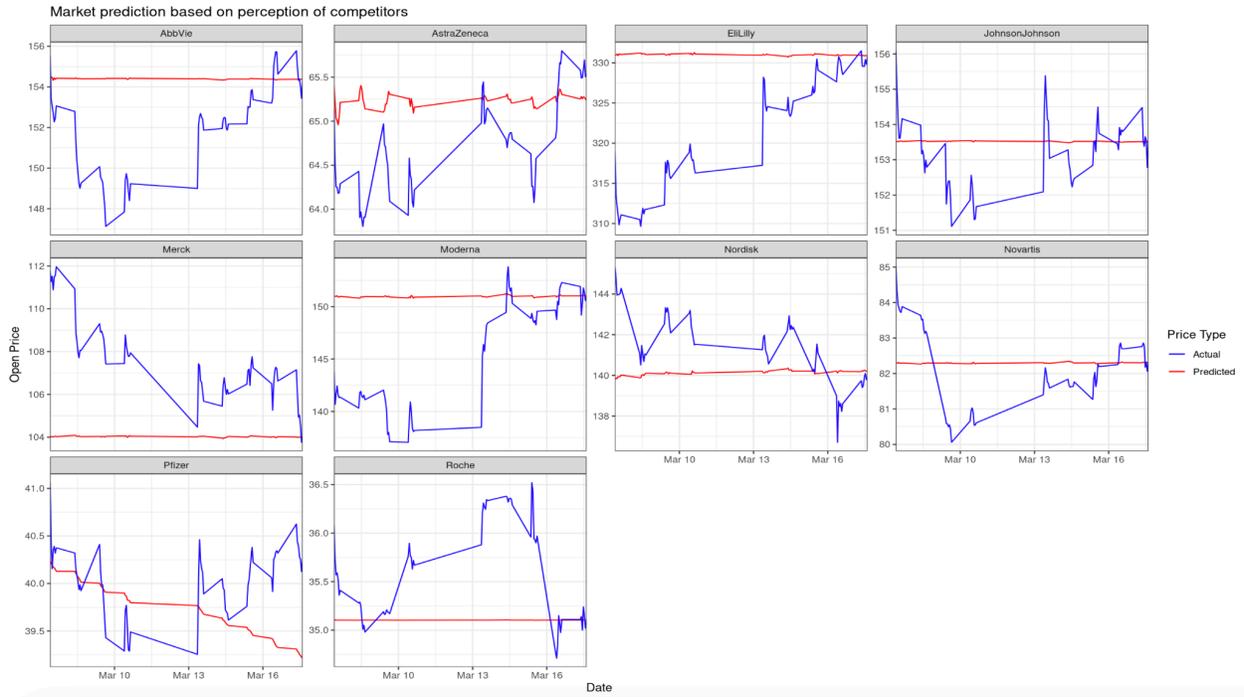

Figure 16: A comparison of the actual and predicted stock market opening prices for each company from March 8th to March 17th, as forecasted by the ARIMA model incorporating historical market opening data and using sentiment analysis proceed data of competitors as covariates.

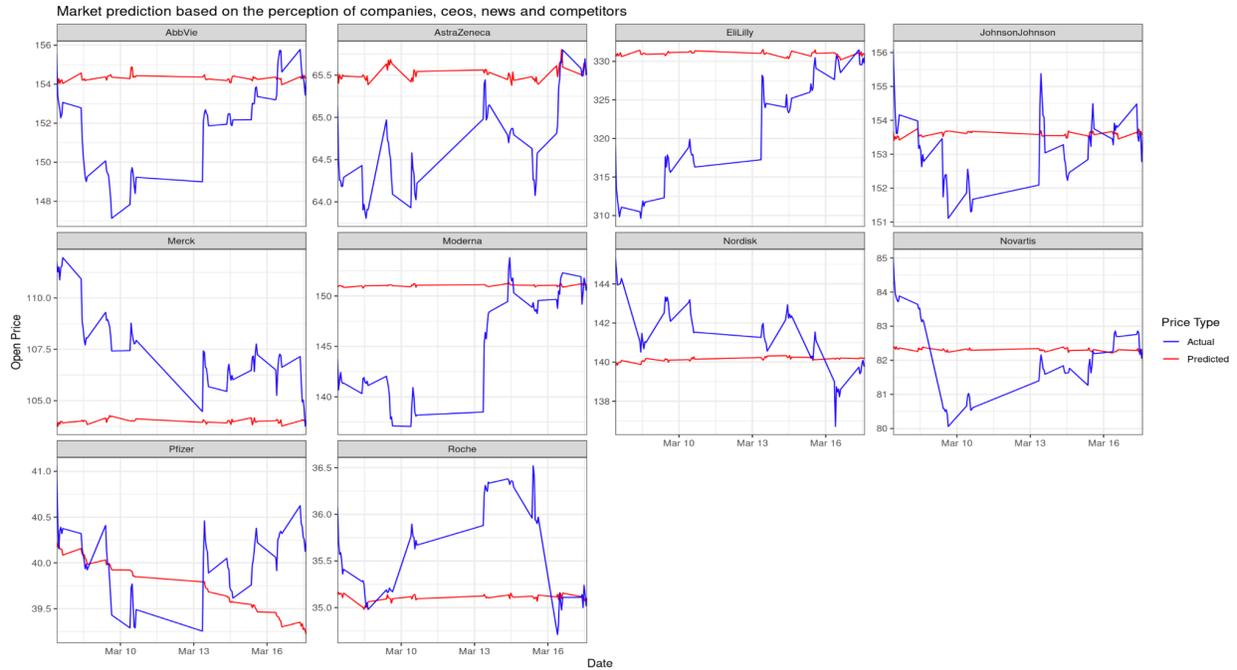

Figure 17: A comparison of the actual and predicted stock market opening prices for each company from March 8th to March 17th, as forecasted by the ARIMA model incorporating historical market opening data and using sentiment analysis of tweets related to the companies, CEOs, COVID-19, vaccines, and competitors simultaneously as covariates. This figure visually demonstrates the accuracy and performance of the ARIMA model in predicting stock market opening prices while considering the influence of various public sentiment factors as covariates in the model, capturing a more comprehensive view of the potential impact of these factors on stock market performance.

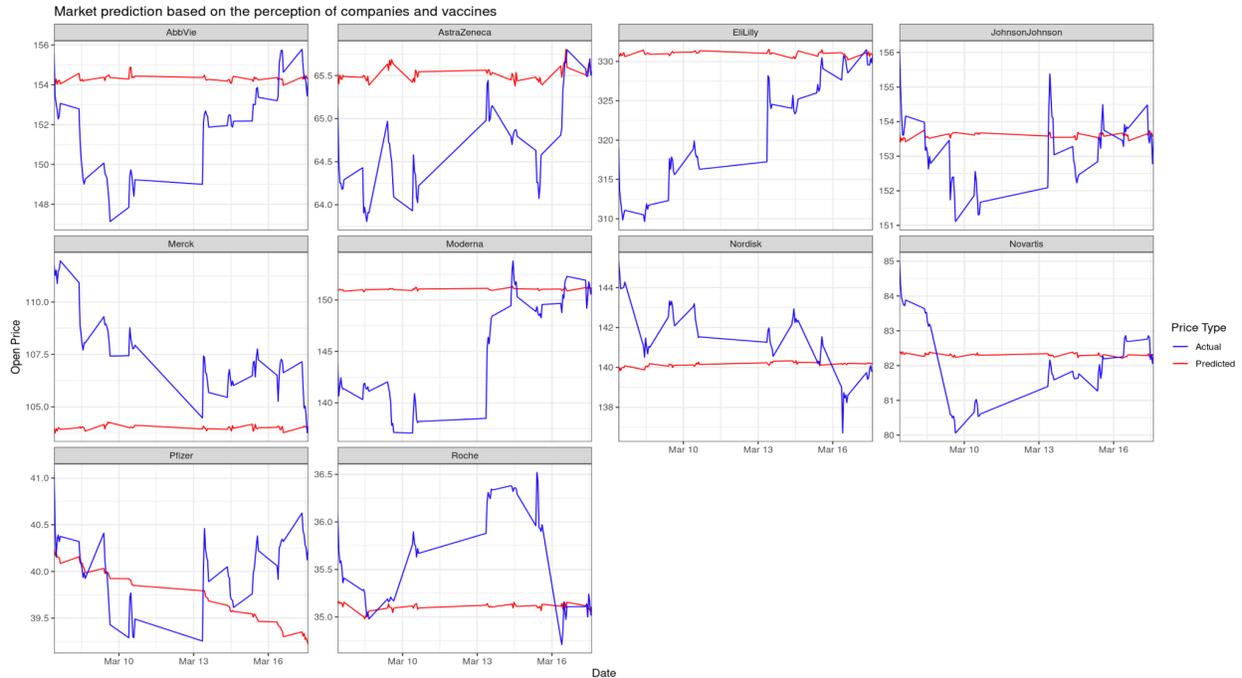

Figure 18: A comparison of the actual and predicted stock market opening prices for each company from March 8th to March 17th, as forecasted by the ARIMA model incorporating historical market opening data and using sentiment analysis of tweets related to both the companies and vaccines as covariates. This figure visually demonstrates the improved accuracy and performance of the ARIMA model in predicting stock market opening prices when considering the combined influence of public sentiment towards companies and vaccines, as these factors have shown better performance in terms of MAPE.

| Company | Hist. record | Companies | CEOs | Vaccine | COVID | Competitors | All | Company&vaccine |
|---|---|---|---|---|---|---|---|---|
| AbbVie | 1.7654779 | 1.7730103 | 1.7760456 | 1.7093467 | 1.768042 | 1.768042 | 1.7384448 | 1.7384448 |
| AstraZeneca | 1.3204773 | 1.3566299 | 1.3191474 | 1.255807 | 1.3218223 | 1.3218223 | 1.2756437 | 1.2756437 |
| EliLilly | 2.9809412 | 2.9297804 | 3.0008533 | 2.9542986 | 2.9959396 | 2.9959396 | 2.9852813 | 2.9852813 |
| Johnson | 0.5255876 | 0.5288334 | 0.5259439 | 0.5301557 | 0.5429978 | 0.5429978 | 0.5457053 | 0.5457053 |
| Merck | 3.2607307 | 3.2248368 | 3.2605173 | 3.2963453 | 3.2569759 | 3.2569759 | 3.2705845 | 3.2705845 |
| Moderna | 4.1593113 | 4.1610435 | 4.1590972 | 4.1470893 | 4.2044133 | 4.2044133 | 4.1617165 | 4.1617165 |
| Nordisk | 1.2530694 | 1.2563138 | 1.2545347 | 1.242369 | 1.2550586 | 1.2550586 | 1.2605519 | 1.2605519 |
| Novartis | 1.1308498 | 1.1309679 | 1.1322648 | 1.1294541 | 1.1249229 | 1.1249229 | 1.1239854 | 1.1239854 |
| Pfizer | 1.1413635 | 1.1413005 | 1.1403317 | 1.1487807 | 1.1253267 | 1.1253267 | 1.1317842 | 1.1317842 |
| Roche | 1.5176051 | 1.524311 | 1.517852 | 1.4979923 | 1.5256422 | 1.5256422 | 1.5092568 | 1.5092568 |
| Mean | 1.9055414 | 1.9027027 | 1.9086588 | 1.8911639 | 1.9121141 | 1.9121141 | 1.9002954 | 1.9002954 |

Table 1: Mean Absolute Percentage Error (MAPE) values for ARIMA predictions of stock market opening prices for each company from March 8th to March 17th, using different sentiment covariates. The table

displays the MAPE values for each company when using historical market opening data alone (Hist. record), and when incorporating sentiment analysis of tweets related to companies, CEOs, vaccines, COVID-19, competitors, all factors combined, and companies & vaccines simultaneously as covariates. The last row of the table shows the mean MAPE value for each covariate across all companies, providing a measure of the overall accuracy and performance of the ARIMA model when considering different public sentiment factors as covariates in predicting stock market opening prices, which we are not overinterpretation in the results.

### 5.6 Forecasting Predictions using Vector Autoregression (VAR) Model

In this last section, we employed the VAR model to predict stock prices for each company in our dataset while simultaneously accounting for the influence of other variables, such as company-specific, COVID-19-related, vaccine-related, CEOs, and competitor sentiments. We believed that the VAR model would be an appropriate choice as it treats each covariate as independent, and we have demonstrated that our covariates are not redundant (Figure 9,10). Using the comprehensive data generated from the preprocessing and sentiment analysis steps, we aimed to analyze the impact of these factors on stock market performance.

Our analysis reveals that in some cases, incorporating the sentiment covariates significantly improved the MAPE (Figure 19-24, Table 2). For instance, AstraZeneca's performance improved considerably when considering the CEO covariate, with a MAPE of 0.9768, much lower than the mean MAPE of 1.6262 for the rest of the categories. Similarly, Nordisk's performance improved when considering the CEO covariate, resulting in a MAPE of 0.9968. These results indicate that the VAR model, which accounts for the influence of various sentiment covariates, can lead to more accurate predictions.

In conclusion, our analysis demonstrates that we successfully predicted stock market performance using the VAR model, which considered the influence of various sentiment factors. By incorporating these covariates, we were able to achieve better predictions in some cases, highlighting the importance of considering diverse factors when modeling and predicting stock prices.

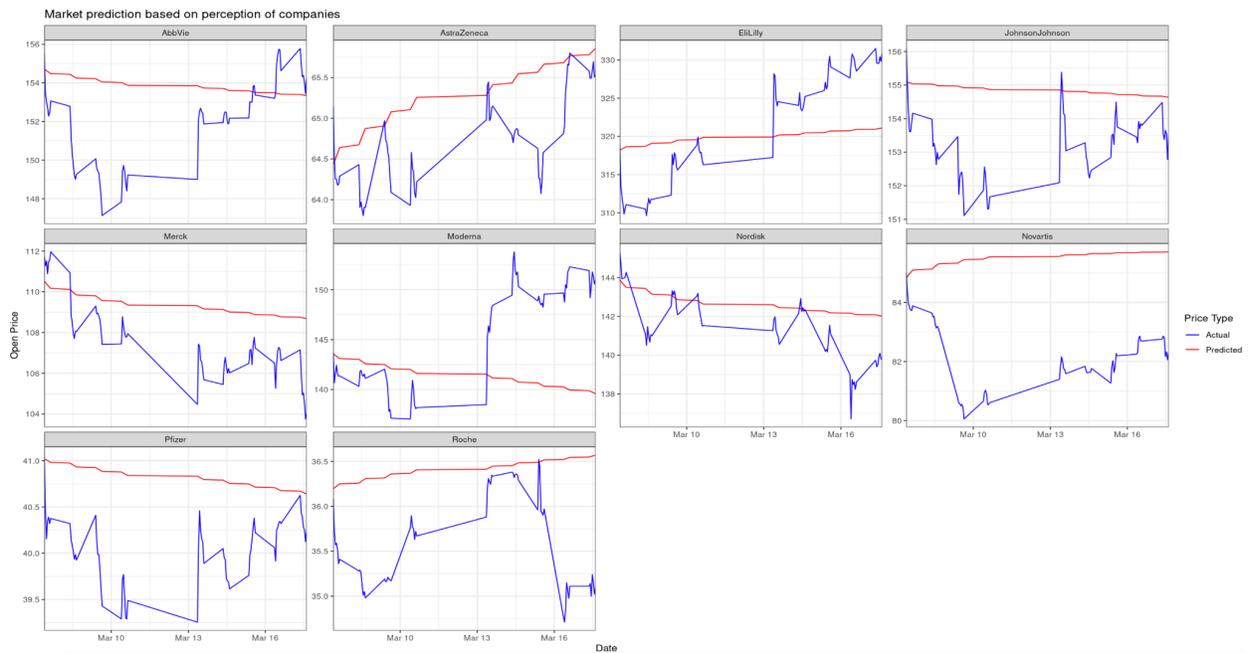

Figure 19: A comparison of the actual and predicted stock market opening prices for each company from March 8th to March 17th, as forecasted by the VAR model incorporating historical market opening data and using sentiment analysis of tweets related to the companies as covariates. This figure visually demonstrates the accuracy and performance of the VAR model in predicting stock market opening prices while considering the influence of public sentiment as covariates in the model.

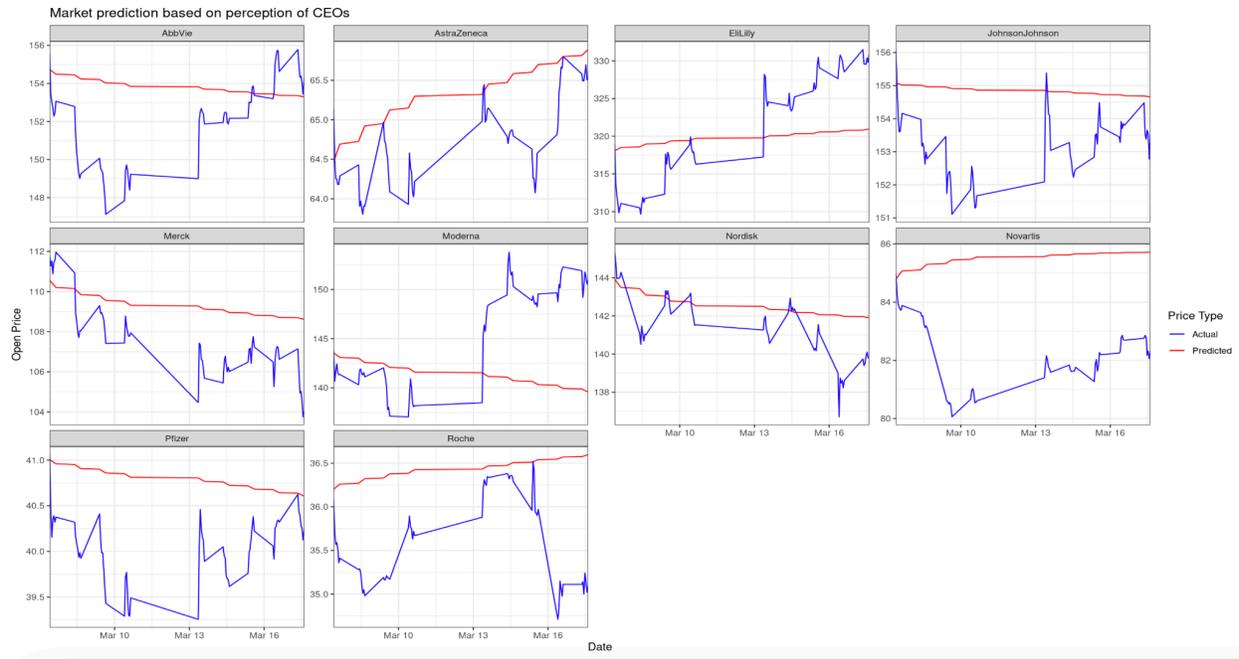

Figure 20: A comparison of the actual and predicted stock market opening prices for each company from March 8th to March 17th, as forecasted by the VAR model incorporating historical market opening data and using sentiment analysis of tweets related to the CEOs as covariates.

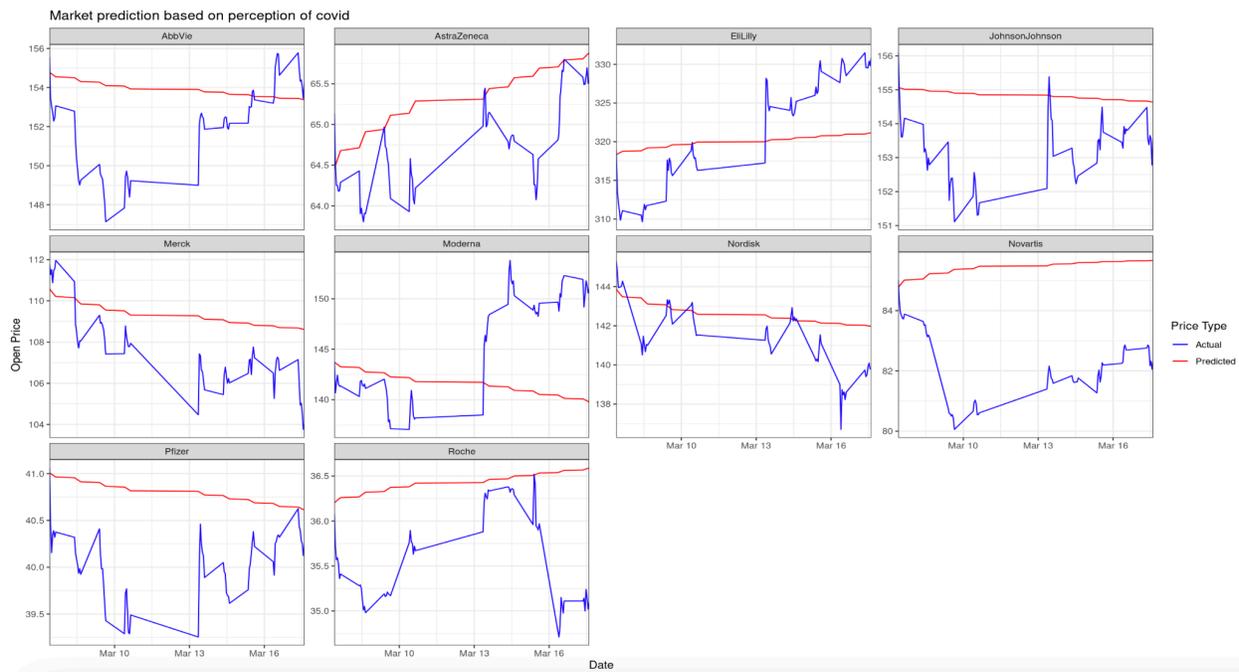

Figure 21: A comparison of the actual and predicted stock market opening prices for each company from March 8th to March 17th, as forecasted by the VAR model incorporating historical market opening data and using sentiment analysis of tweets related to COVID as covariates.

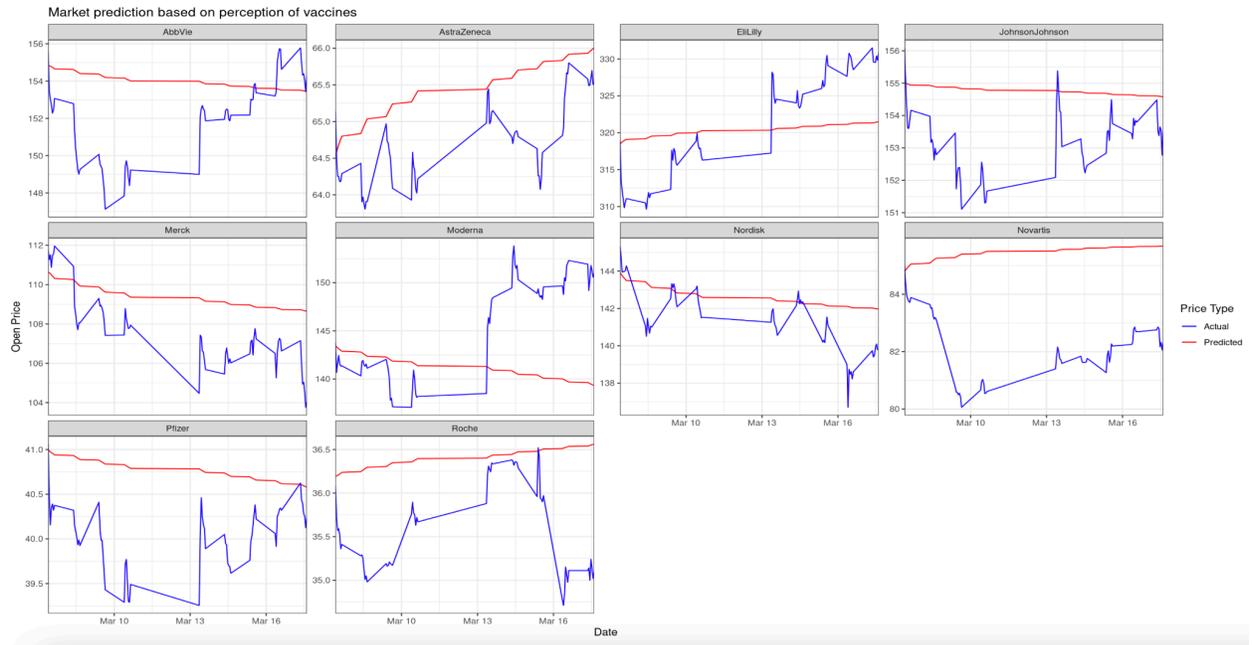

Figure 22: A comparison of the actual and predicted stock market opening prices for each company from March 8th to March 17th, as forecasted by the VAR model incorporating historical market opening data and using sentiment analysis of tweets related to the vaccines as covariates.

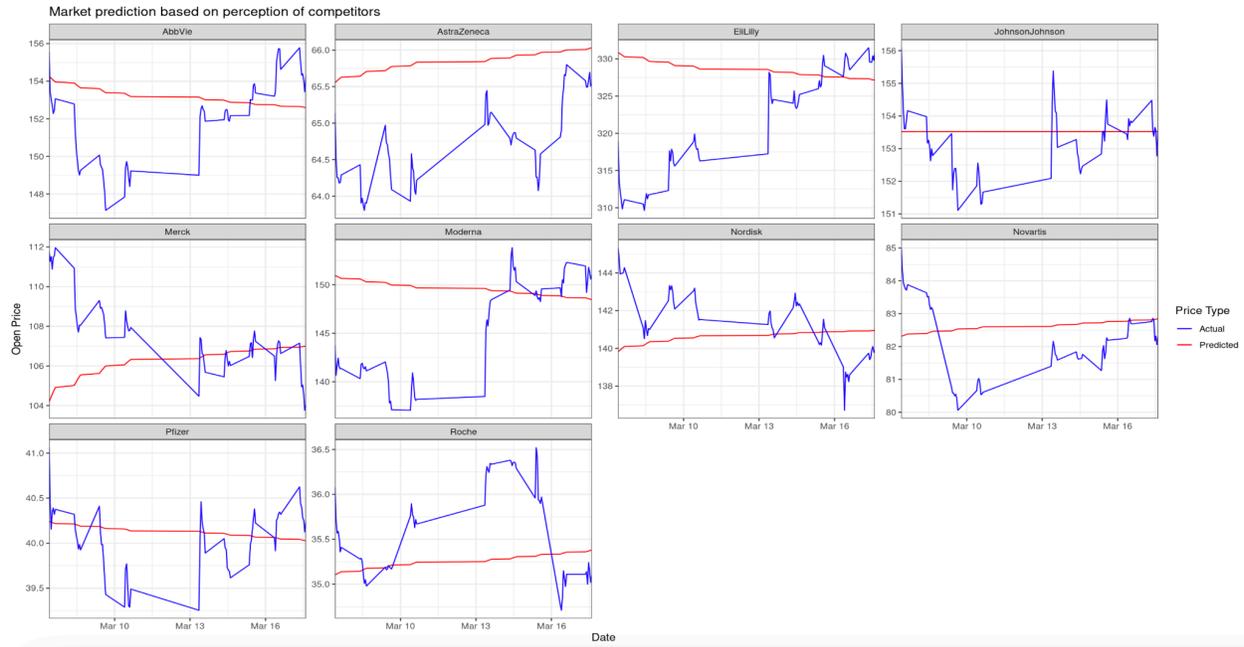

Figure 23: A comparison of the actual and predicted stock market opening prices for each company from March 8th to March 17th, as forecasted by the VAR model incorporating historical market opening data and using sentiment analysis of tweets related to the competitors as covariates.

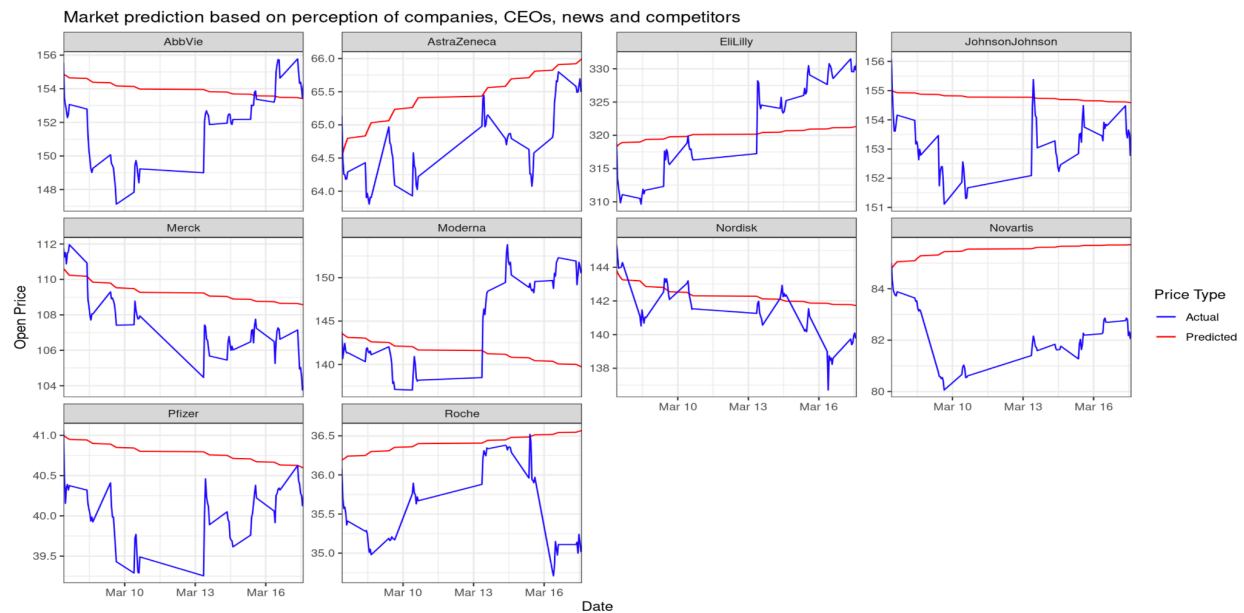

Figure 24: A comparison of the actual and predicted stock market opening prices for each company from March 8th to March 17th, as forecasted by the VAR model incorporating historical market opening data

and using sentiment analysis of tweets related to the companies, CEOs, COVID-19, vaccines, and competitors simultaneously as covariates. This figure visually demonstrates the accuracy and performance of the VAR model in predicting stock market opening prices while considering the influence of various public sentiment factors as covariates in the model, capturing a more comprehensive view of the potential impact of these factors on stock market performance.

| Company | Companies | CEOs | Vaccine | COVID | Competitors | All |
|---------|-----------|------|---------|-------|-------------|-----|
| AbbVie | 1.3397776 | 1.6427819 | 1.701782 | 1.6673301 | 1.4773286 | 1.6937921 |
| AstraZeneca | 1.9151156 | 0.9768085 | 1.143815 | 0.9641481 | 1.7474646 | 1.1332647 |
| EliLilly | 2.4082974 | 1.8348834 | 1.821239 | 1.8313728 | 2.6460867 | 1.8247675 |
| Johnson | 0.5291664 | 1.0962024 | 1.047234 | 1.0874922 | 0.525663 | 1.0466666 |
| Merck | 1.6485757 | 1.9233214 | 1.949397 | 1.9165676 | 1.9381894 | 1.8860731 |
| Moderna | 3.760676 | 4.0938909 | 4.115048 | 4.0775476 | 3.9498702 | 4.0686873 |
| Nordisk | 1.1126908 | 0.9968463 | 1.018771 | 1.0218158 | 1.2167739 | 0.9393601 |
| Novartis | 1.4416417 | 4.1312918 | 4.079591 | 4.0516057 | 1.3281317 | 4.12 |
| Pfizer | 0.7356179 | 1.8822464 | 1.827206 | 1.8940481 | 0.7260857 | 1.8573541 |
| Roche | 1.3706057 | 2.3656277 | 2.285367 | 2.3539372 | 1.4212828 | 2.2991885 |
| Mean | 1.6262165 | 2.0943901 | 2.098945 | 2.0865865 | 1.6976876 | 2.0869154 |

Table 2: Mean Absolute Percentage Error (MAPE) values for VAR predictions of stock market opening prices for each company from March 8th to March 17th, using different sentiment covariates. The table displays the MAPE values for each company when incorporating sentiment analysis of tweets related to companies, CEOs, vaccines, COVID-19, competitors, and all factors combined as covariates in the VAR model. The last row of the table shows the mean MAPE value for each covariate across all companies, providing a measure of the overall accuracy and performance of the VAR model when considering different public sentiment factors as covariates in predicting stock market opening prices, which we are not overinterpreting in the results.

## 6. CONCLUSION

Our study aimed to investigate the relationship between news sentiment, biotech companies, their CEOs, and stock market performance, using a combination of time series forecasting techniques and sentiment analysis. Through the analysis of historical stock market data, we identified increased volatility in stock prices during after-hours trading. Our Twitter data analysis revealed the public's engagement with biotech companies, CEOs, COVID-19, and vaccine-related news, with some companies and CEOs receiving more attention, potentially impacting their stock market performance.

The VADER sentiment analysis tool allowed us to explore public sentiment patterns related to these topics. While some companies and CEOs received more favorable public opinions, certain subjects like vaccines tended to provoke more polarized reactions. The analysis of relationships between sentiment scores showed a significant positive correlation between a company's sentiment scores and its competitors', highlighting the importance of considering competitor sentiment scores when analyzing factors affecting stock prices

Our forecasting predictions using ARIMA and VAR models demonstrated that incorporating news sentiment covariates into the models led to improved predictions of stock market performance. The results emphasize the importance of considering various factors, such as company news, CEO news, vaccine news, COVID news, and competitor news when modeling and predicting stock prices. This study provides valuable insights into the complex relationship between news sentiment, biotech companies, their CEOs, and stock market performance, and paves the way for future research incorporating diverse factors for more accurate stock price predictions.

## 7. PERSPECTIVES

While our study has provided valuable insights into the complex relationship between news sentiment, biotech companies, their CEOs, and stock market performance, there are several potential directions for future work to further enhance our understanding and improve the accuracy of stock price predictions. In future research, we could expand the scope of the study by considering a larger and more diverse dataset, including more biotech companies, a broader range of industries, and additional news sources such as news articles, blogs, and forums. This would provide a more comprehensive picture of public sentiment and its impact on stock market performance across various sectors.

Future work could delve deeper into the impact of specific events or announcements, such as product approvals, clinical trial results, or financial reports, on stock market performance. By analyzing these events and their corresponding sentiment, we could better understand how specific news events influence stock prices and public sentiment. Although the VADER sentiment analysis tool performed well in our study, future research could explore other sentiment analysis techniques, such as machine learning models or deep learning techniques like LSTM and BERT. These methods may provide more accurate sentiment scores and help to uncover more nuanced sentiment patterns in the collected data.

To improve the accuracy of stock price predictions, future work could investigate the inclusion of additional features and covariates, such as macroeconomic indicators, technical analysis, or sector-specific factors. This would provide a more comprehensive model, accounting for a broader range of factors influencing stock market performance. In our study, we employed the ARIMA and VAR models for forecasting stock prices. Future research could compare the performance of these models with other time series forecasting techniques, such as state space models, Bayesian structural time series, or machine learning-based models like LSTM (Long Short-Term Memory) and GRU (Gated Recurrent Unit). This comparison would help identify the most suitable model for predicting stock prices in the context of news sentiment and other factors.

One promising direction for future work is the development of real-time sentiment analysis and stock price prediction systems. These systems would continuously collect and analyze news sentiment data, providing real-time insights into the potential impact of news events on stock prices. Such systems could be valuable tools for investors, enabling them to make more informed decisions and better manage risk. By pursuing these research directions, we can continue to advance our understanding of the relationship between news sentiment, biotech companies, their CEOs, and stock market performance, and develop more effective and accurate stock price prediction models.

## 8. SUPPLEMENTARY

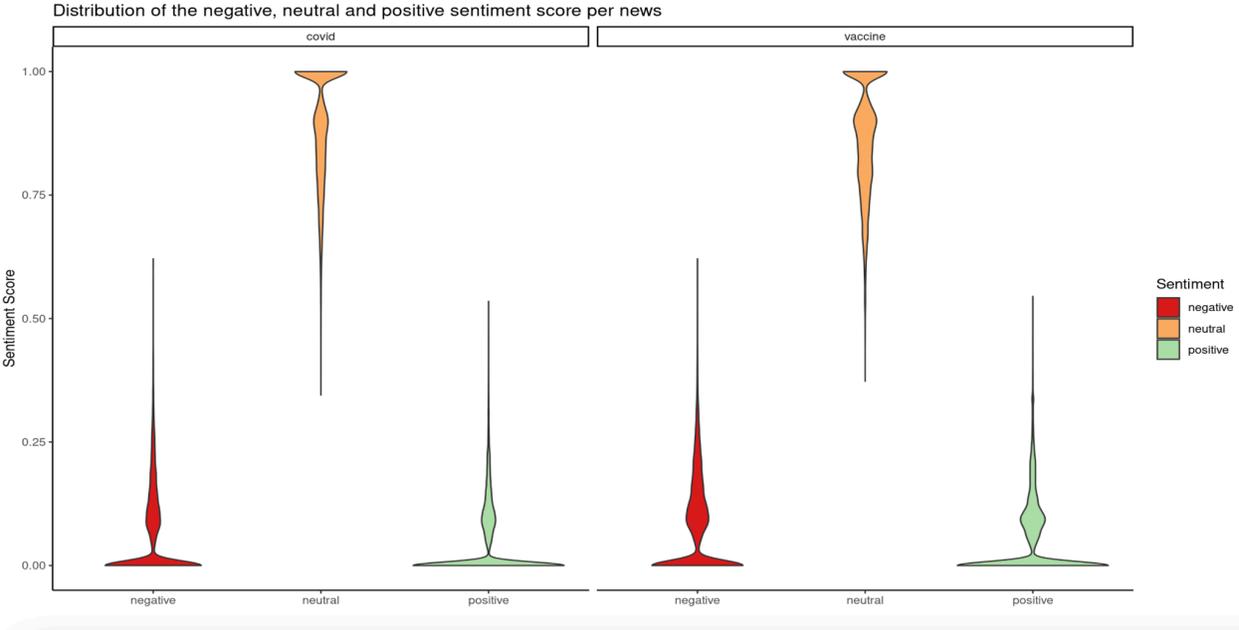

Figure S1: Distribution of negative, neutral, and positive sentiment scores obtained from the sentiment analysis of tweets related to CEOs.

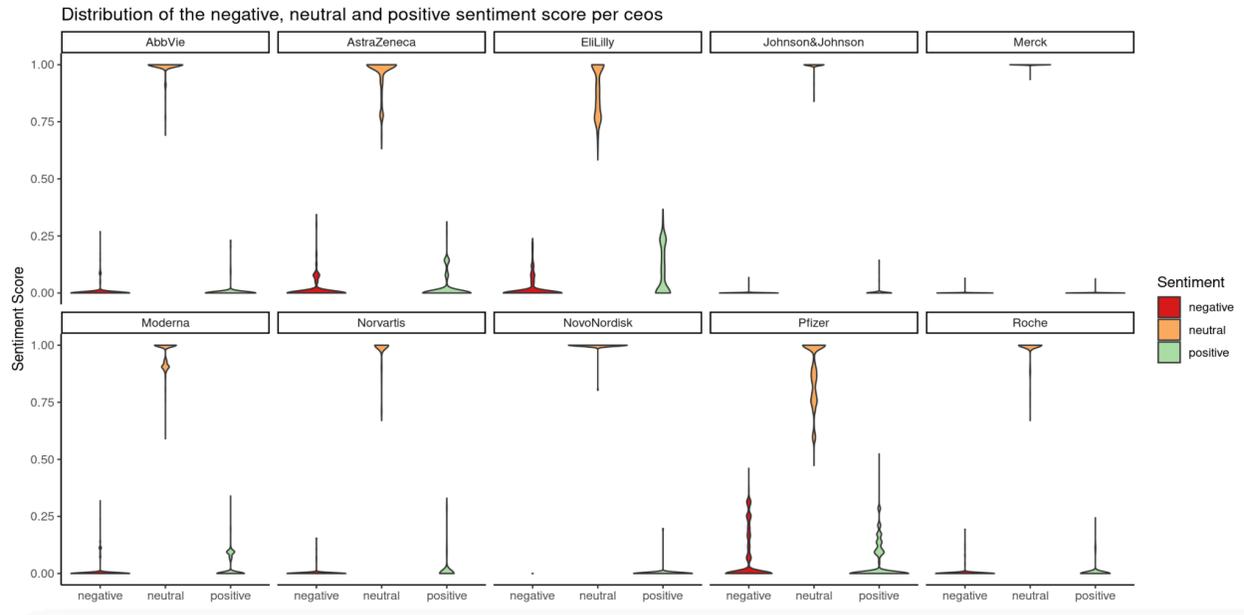

Figure S2: Distribution of negative, neutral, and positive sentiment scores obtained from the sentiment analysis of tweets related to the news which include COVID and vaccines.